%% file: paper.tex
\documentclass[conference]{IEEEtran}
\IEEEoverridecommandlockouts
\newcommand{\shadd}[1]{\textcolor{blue}{#1}}
\makeatother
\newtheorem{example}{Example}
\usepackage{cite}
\usepackage{amsmath,amssymb,amsfonts}
\usepackage{algorithmic}
\usepackage{graphicx}
\usepackage{textcomp}
\usepackage{xcolor}
\def\BibTeX{{\rm B\kern-.05em{\sc i\kern-.025em b}\kern-.08em
    T\kern-.1667em\lower.7ex\hbox{E}\kern-.125emX}}
\usepackage{enumitem}
\usepackage{cleveref}
\usepackage{pifont}
\definecolor{LightCyan}{rgb}{0.88,1,1}
\definecolor{LightRed}{rgb}{1,0.88,1}
\definecolor{LightYellow}{rgb}{1,1,0.88}
\definecolor{LightGray}{gray}{0.8}
\input{custom_sytles}

\input{command}

\begin{document}

\title{Community Search: A Meta-Learning Approach}

\author{%
	{Shuheng Fang\IEEEauthorrefmark{1}, Kangfei Zhao\thanks{\IEEEauthorrefmark{2} Corresponding author.}\IEEEauthorrefmark{2}, Guanghua Li\IEEEauthorrefmark{3}, Jeffrey Xu Yu\IEEEauthorrefmark{1}} 
	\vspace{1.6mm}\\
	\fontsize{10}{10}\selectfont\itshape
	\IEEEauthorrefmark{1}The Chinese University of Hong Kong,
	\IEEEauthorrefmark{2}Beijing Institute of Technology\\
	\IEEEauthorrefmark{3}The Hong Kong University of Science and Technology (Guangzhou)\\
	
	\fontsize{9}{9}\selectfont\ttfamily\upshape
	
	\IEEEauthorrefmark{1}{\{shfang,yu\}}@se.cuhk.edu.hk, \IEEEauthorrefmark{2}zkf1105@gmail.com,
	\IEEEauthorrefmark{3}gli945@connect.hkust-gz.edu.cn
	\fontsize{9}{9}\selectfont\ttfamily\upshape
	\vspace{-0.6cm}
}
\vspace{-0.6cm}

\maketitle
\thispagestyle{plain}

\comment{
\author{\IEEEauthorblockN{Shuheng Fang}
\IEEEauthorblockA{\textit{The Chinese University of Hong Kong} \\
shfang@se.cuhk.edu.hk}

\and
\IEEEauthorblockN{Kangfei Zhao}
\IEEEauthorblockA{\textit{Beijing Institute of Technology} \\
zkf1105@gmail.com}

\and
\IEEEauthorblockN{Guanghua Li}
\IEEEauthorblockA{\textit{Wuhan University} \\
guanghli@whu.edu.cn}

\and
\IEEEauthorblockN{Jeffrey Xu Yu}
\IEEEauthorblockA{\textit{The Chinese University of Hong Kong} \\
yu@se.cuhk.edu.hk}
}
\maketitle}

\begin{abstract}
Community Search (CS) is one of the fundamental graph analysis tasks, which is a building block of various real applications.
Given any query nodes, CS aims to find cohesive subgraphs that query nodes belong to. 
Recently,  a large number of CS algorithms are designed. 
These algorithms adopt pre-defined subgraph patterns to model the communities, which cannot find ground-truth communities that do not have such pre-defined patterns in real-world graphs.
Thereby, machine learning (ML) and deep learning (DL) based approaches are proposed to capture flexible community structures by learning from ground-truth communities in a data-driven fashion.
These approaches rely on sufficient training data to provide enough generalization for ML models, however, the ground-truth cannot be comprehensively collected beforehand.

In this paper, we study ML/DL-based approaches for CS, under the
circumstance of small training data.  Instead of directly fitting the
small data, we extract prior knowledge which is shared across multiple
CS tasks via learning a meta model. Each CS task is a graph with
several queries that possess corresponding partial ground-truth.  The
meta model can be swiftly adapted to a task to be predicted by feeding
a few task-specific training data.  We find that trivially applying multiple
classical meta-learning algorithms to CS suffers from problems
regarding prediction effectiveness, generalization capability and
efficiency.  To address such problems, we propose a novel
meta-learning based framework, Conditional Graph Neural Process
(CGNP), to fulfill the prior extraction and adaptation procedure.  A
meta CGNP model is a task-common node embedding function for
clustering, learned by metric-based graph learning, which
fully exploits the characteristics of CS.  We compare CGNP with
CS algorithms and ML baselines on real graphs with
ground-truth communities.
Our experiments verify that CGNP outperforms the other native
graph algorithms and ML/DL baselines 0.33 and 0.26 on \Fone score by average.
The source code has been made available at
\textbf{https://github.com/FangShuheng/CGNP}.
\end{abstract}


\section{Introduction}

\begin{figure}[t] 
	\centering 
	\includegraphics[width=0.34\textwidth]{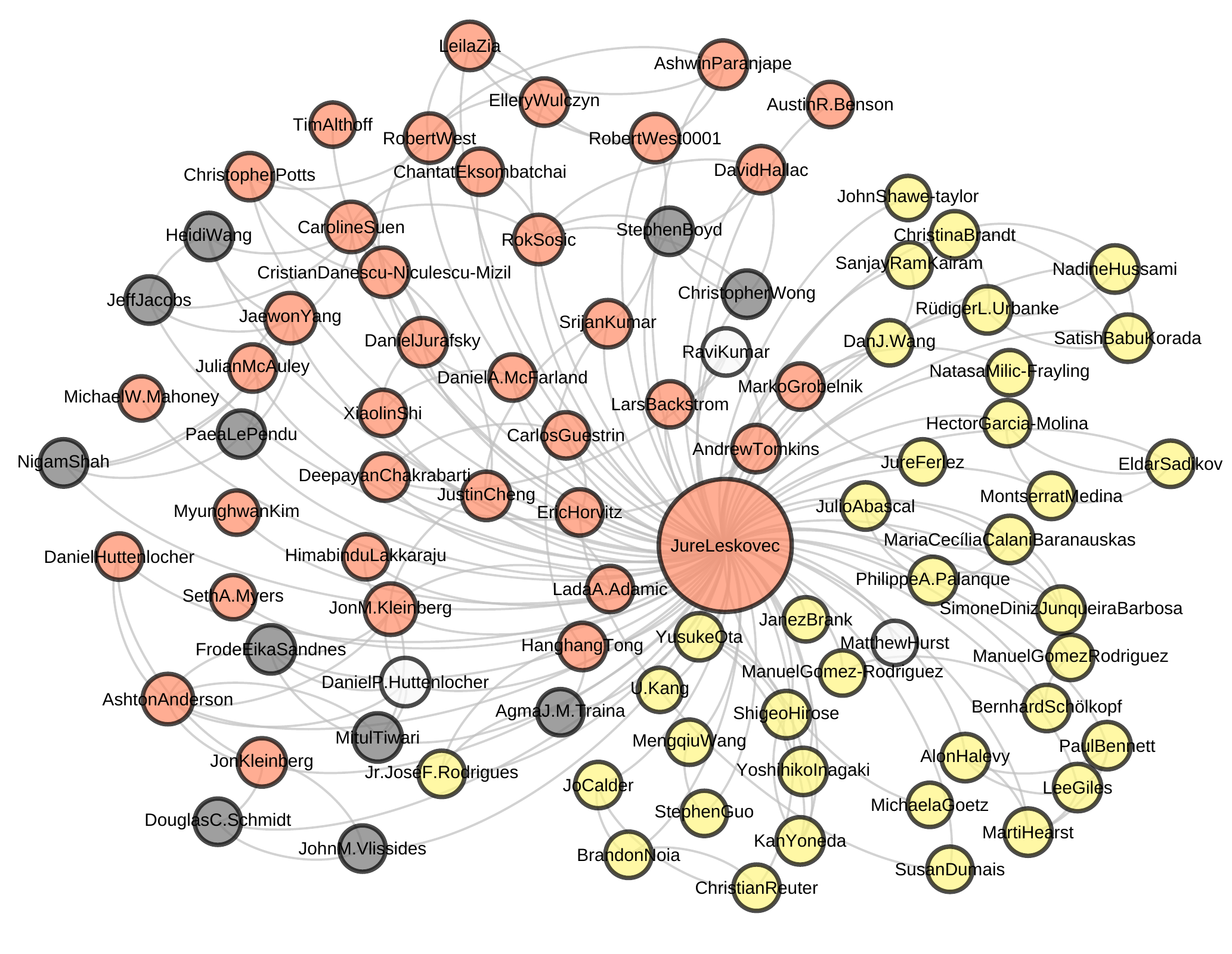} 
	\vspace{-0.4cm}
	\caption{An Example on DBLP: Query `Jure Leskovec'} 
	\vspace{-0.4cm}
	\label{fig:case} 
\end{figure}

Community is a cohesive subgraph that is densely intra-connected and
loosely inter-connected in a graph. Given any query nodes, community
search (CS) aims at finding communities covering the query nodes,
i.e., local query-dependent communities, which has a wide range of
real applications, e.g., friend recommendation, advertisement in
e-commence and protein complex
identification~\cite{DBLP:journals/vldb/FangHQZZCL20,
  DBLP:series/synthesis/2019Huang}.
%
%
In the literature, to model structural cohesiveness, various
community models are adopted, including $k$-core~\cite{cs3,cs4,cs6},
$k$-truss~\cite{cs2,cs7}, $k$-clique \cite{cs1,cs8} and $k$-edge
connected component~\cite{cs9,cs10}. Such models can be computed
efficiently by CS algorithms.
But such models are designed based on some pre-defined community
patterns which are too rigid to be used to find ground-truth
communities in real applications.  We show a DBLP example in
Fig.~\ref{fig:case} in which nodes represent researchers
%
%
and edges represent their collaboration.
%
%
The ground-truth community of 'Jure Leskovec', i.e., the orange and
white nodes in Fig.~\ref{fig:case} are with the researchers who have
collaborations and share the common interest of 'social
networks'. Such a community cannot be accurately found with any
$k$-related subgraph patterns. For example, in the community, some
nodes (e.g., Michael W. M.) have one neighbor, which can only be found
by $1$-core that may result in accommodating the whole graph.

\comment{
To this end,
it is quite difficult to choose a proper $k$ value as well as one
community metric to pursue high accuracy.

In other words, in real-world
graphs, it is difficult to have some universal community constraints
for us to

which is the overall best. One constraint may be either too loose or
too tight.

Even for one graph, the topology is diverse in different
subgraphs so that a fixed community constraint may not be consistently
applicable for different local queries.
}

To tackle the structural inflexibility of CS algorithms, ML/DL-based
solutions~\cite{ICSGNN, AQDGNN} are arising as an attractive research
direction.  They build ML/DL models from given ground-truth
communities and expect the models to generalize to unknown
community-member relationships.
Such ML/DL-based approaches have achieved success in finding
high-quality solutions due to two reasons.  For one thing, these
data-driven approaches get rid of the inflexible constraints and adapt
to implicit structural patterns from data. For another thing, the
models can learn via error feedback from its predictions on the query
nodes in the ground-truth communities.
%
%
But, effective error feedback heavily relies on sufficient
ground-truth communities to train, which are hard to collect and
label. On the one hand, they are labor-intensive, on the other hand, such
ground-truth communities for different query nodes can be very
different.

\comment{
Specifically, \cite{ICSGNN}
collects user feedback to update a model incrementally and
interactively. The model is trained for a specific query node, whose
effectiveness is fully determined by the quality of the given
ground-truth of that query node.
%
%
\cite{AQDGNN} proposes a graph neural network based model that is
trained by a collection of query nodes with their ground-truth, and
makes predictions for unseen query nodes.  For one graph, a large
volume of query nodes with ground-truth communities are necessary for
training, which ensure the model well generalizes to other local
queries.
}

To deal with this problem, an effective solution is to inject prior
knowledge extracted from multiple CS tasks into the ML model,
%
%
where one CS task is a subgraph with a small number of
query nodes with partial ground-truth community membership.
The implicit prior knowledge of the CS tasks is rather intuitive,
i.e., for any query node of an arbitrary graph, its communities are
the nearby densely connected nodes that share similar attributes with
the query node. Such prior is shared by different CS tasks for
different query nodes in any real-world graphs.
%
%
%
%
%
\comment{
For example, the meta-model is constructed by multiple tasks, which
are ego-networks of researchers in different fields, possessing
several query nodes with ground-truth as training data.  Then, given
several query nodes, e.g., `U. Kang', with partial ground-truth for
the adaptation, the model can be adapted to the task of
Fig.~\ref{fig:case} to search communities for new query nodes, e.g.,
`Jure'.
}
And the prior knowledge is capable of synthesizing similar or
complementary inductive bias across different CS tasks to compensate
the insufficient knowledge from small training data, thus can be
swiftly adapted to a new task to test.
In this paper, we concentrate ourselves on learning a meta model to
capture this prior by meta-learning.

There are existing meta-learning algorithms, e.g., simple feature
transfer and model-agnostic meta-learning. However, trivial
adaptations to CS tasks fail to achieve high performance since they do
not exploit the intrinsic characteristic of the CS tasks.  For CS,
what a model needs to justify for each node in a graph is whether or
not it has its community membership with any given query node.  To
facilitate such binary justification, we propose a novel model,
Conditional Graph Neural Process (CGNP), to generate node embeddings
conditioned on the small training data, where the distance between a
node embedding to that of the query node explicitly indicates their
community membership.  Furthermore, as a graph specification of
Conditional Neural Process (CNP)~\cite{CNP}, CGNP inherits the main
ideas of CNP that implicitly learns a kernel function between a
training query node and a query node to be predicted.  \comment{ In a
  nutshell, our learned CGNP is a common embedding function that
  transforms the nodes of every graph into a distance-aware hidden
  space, as well as a common kernel function between query nodes
  across different graphs, which captures the CS prior knowledge.  }
In a nutshell, the learned CGNP is not only a \emph{common embedding
  function} but also a \emph{common kernel function}, shared across
different graphs. The embedding function transforms the nodes of each
graph into a distance-aware hidden space, while the kernel function
memorizes the small training data of each task as a hidden
representation.  Compared with optimization-based meta-learning
approaches whose parameters are easy to overfit, the metric learning
and memorization mechanisms are more suitable for classification tasks
with small training data, especially for imbalanced labels.
%
%
%

The contributions of this paper are summarized as follows:
%
\ding{172} We formulate the problem of learning a meta model to answer CS queries, where the meta model is to absorb the prior knowledge from multiple CS tasks and adapt to a specific task with only a few training data.  We generalize three CS task scenarios that represent comprehensive query cases. To the best of our knowledge, our study is the first attempt at meta model/algorithm for CS.
\ding{173} We explore three Graph Neural Network based solutions, i.e., feature transfer, model-agnostic meta-learning and Graph Prototypical Network, which are trivial adaptations of existing transfer/meta-learning algorithms to CS. 
We identify their individual limitations regarding prediction effectiveness, generalization capability and efficiency.
\ding{174} We propose a novel framework, \emph{Conditional Graph Neural Process} (CGNP) on the basis of conceptual CNP and learn the meta model in an efficient, metric-based learning perspective. We design and explore model variants with different model complexities and different options for the core components. To the best of our knowledge, we made the first effort to explore how to solve CS problem by meta-learning.
%
\ding{175} We conduct extensive experiments on 6 real-world datasets
with ground-truth communities for performance evaluation. Compared
with 3 CS algorithms, 4 naive approaches, and 3 supervised
learning validates our CGNP outperforms the others
with small training and prediction cost.

\stitle{Roadmap:} 
The rest of the paper is organized as follows. 
\cref{sec:related} reviews the relative work. In \cref{sec:problem}, we give the problem statement followed by three naive solutions introduced in \cref{sec:naive}.
We introduce the core idea of our approach, CGNP in \cref{sec:metric}. We elaborate on its architecture design and present the learning algorithms of CGNP in \cref{sec:CGNP}. 
We present our comprehensive experimental studies in \cref{sec:exp} and conclude the paper in \cref{sec:conclusion}.

\input{related}

\input{main}

\input{experiment}

\section{Conclusion}
\label{sec:conclusion}
In this paper, we study leveraging ML/DL approaches for community search (CS), under the circumstance that the training data is scarce. 
We propose a metric-based meta-learning framework, Conditional Graph Neural Process (CGNP) to learn a meta model to capture the prior knowledge of CS.
The meta model is adapted to a new task swiftly to make predictions of the community membership, where a task is a graph with only a few given ground-truth. 
To the best of our knowledge, CGNP is the first meta-learning model for CS that utilizes the generalization ability of neural networks to the greatest extent.
Compared with algorithmic approaches, CGNP supports flexible community structures learned from the data. Compared with general meta-learning algorithms, CGNP further exploits the characteristic of CS. Our extensive experiments demonstrate that CGNP outperforms the two lines of approaches significantly regarding accuracy and efficiency.

\section*{Acknowledgment}
This work was supported by the Research Grants Council of Hong Kong, China, No. 14203618, No. 14202919 and No. 14205520.

\bibliographystyle{IEEEtran}
\bibliography{ref}

\end{document}

%% file: custom_sytles.tex
\usepackage{graphicx}
\usepackage{balance}  
\usepackage[linesnumbered,vlined,ruled]{algorithm2e}
\usepackage{lipsum}
\usepackage{amsmath}
\usepackage{physics}
\usepackage{subfigure}
\usepackage{tabularx}
\usepackage{color}
\usepackage{colortbl}
\usepackage{float}
\usepackage{listings}
\usepackage{multirow}
\usepackage[normalem]{ulem}
 \usepackage{longtable}
\usepackage{enumitem}
\usepackage{multirow}
\usepackage{booktabs}
\usepackage{capt-of}
\usepackage{pifont}
\usepackage{ulem}
\usepackage{soul}
\usepackage{cancel}
\usepackage{bm}
\usepackage{url}
\usepackage{caption}
\usepackage{tikz}

\def\BibTeX{{\rm B\kern-.05em{\sc i\kern-.025em b}\kern-.08em
    T\kern-.1667em\lower.7ex\hbox{E}\kern-.125emX}}

 \graphicspath{{./Graph/}, {./Fig/}, {./Legend/}}

\long\def\comment#1{}

\newcounter{definition}[section]
\renewcommand{\thedefinition}{\nthesection.\arabic{definition}}

\newcounter{theorem}[section]
\renewcommand{\thetheorem}{\nthesection.\arabic{theorem}}

\newcounter{lemma}[section]
\renewcommand{\thelemma}{\nthesection.\arabic{lemma}}
        
\newcounter{proposition}[section]
\renewcommand{\theproposition}{\nthesection.\arabic{proposition}}

\newcounter{remark}[section]
\renewcommand{\theremark}{\nthesection.\arabic{remark}}

\newcommand{\nthesection}{\arabic{section}}

\newcommand{\stitle}[1]{\vspace{1ex} \noindent{\bf #1}}

\newcommand{\green}[1]{\textcolor{green}{}}

\newcommand{\kw}[1]{{\ensuremath {\mathsf{#1}}}\xspace}

%% file: command.tex
\newcommand{\etitle}[1]{\vspace{1ex} \noindent{\underline{\em #1}}}

\newcommand{\model} {\mathcal{M}}
\newcommand{\task} {\mathcal{T}}
\newcommand{\query} {\mathcal{Q}}
\newcommand{\support} {\mathcal{S}}
\newcommand{\community} {\mathcal{C}}
\newcommand{\neighbor} {\mathcal{N}}

\newcommand{\Fagg}{f_{\mathcal{A}}}
\newcommand{\Fcom}{f_{\mathcal{C}}}
\newcommand{\loss}{\mathcal{L}}
\newcommand{\relu}{\mathsf{ReLU}}
\newcommand{\sigmoid}{\mathsf{sigmoid}}
\newcommand{\softmax}{\mathsf{softmax}}

\newcommand{\Reddit} {\kw{Reddit}}
\newcommand{\Cora} {\kw{Cora}}
\newcommand{\Citeseer} {\kw{Citeseer}}
\newcommand{\Arxiv} {\kw{Arxiv}}
\newcommand{\DBLP} {\kw{DBLP}}
\newcommand{\Facebook} {\kw{Facebook}}
\newcommand{\Coraciteseer} {\kw{Cora2Cite}}
\newcommand{\Citeseercora} {\kw{Cite2Cora}}

\newcommand{\Fone} {\kw{F1}}
\newcommand{\Pre} {\kw{Pre}}
\newcommand{\Rec} {\kw{Rec}}
\newcommand{\Acc} {\kw{Acc}}
\newcommand{\distance} {\kw{dist}}
\newcommand{\Kernel} {\mathcal{K}_\theta}

\newcommand{\SGSC}{{$\texttt{SGSC}$}\xspace}
\newcommand{\SGDC}{{$\texttt{SGDC}$}\xspace}
\newcommand{\MGOD}{{$\texttt{MGOD}$}\xspace}
\newcommand{\MGDD}{{$\texttt{MGDD}$}\xspace}

\newcommand{\ATC}{{$\texttt{ATC}$}\xspace}
\newcommand{\ACQ}{{$\texttt{ACQ}$}\xspace}
\newcommand{\Featrans}{{$\texttt{FeatTrans}$}\xspace}
\newcommand{\MAML}{{$\texttt{MAML}$}\xspace}
\newcommand{\Supervise}{{$\texttt{Supervised}$}\xspace}
\newcommand{\CGNPIP}{{$\texttt{CGNP-IP}$}\xspace}
\newcommand{\CTC}{{$\texttt{CTC}$}\xspace}
\newcommand{\CGNPMLP}{{$\texttt{CGNP-MLP}$}\xspace}
\newcommand{\CGNPGNN}{{$\texttt{CGNP-GNN}$}\xspace}
\newcommand{\ICSGNN}{{$\texttt{ICS-GNN}$}\xspace}
\newcommand{\PN}{{$\texttt{GPN}$}\xspace}
\newcommand{\Reptile}{{$\texttt{Reptile}$}\xspace}
\newcommand{\AQDGNN}{{$\texttt{AQD-GNN}$}\xspace}

\newcommand{\GCN}{{\sl GCN}\xspace}

\newcommand{\GAT}{{\sl GAT}\xspace}
\newcommand{\SAGE}{{\sl GraphSAGE}\xspace}

%% file: related.tex
\section{Related Work}
\label{sec:related}

\stitle{Community Search.}
A comprehensive survey of CS problems and approaches can be found in~\cite{DBLP:journals/vldb/FangHQZZCL20, DBLP:series/synthesis/2019Huang}. 
In a nutshell, CS problem can be divided into two categories. One is non-attributed community search which only concerns the structural cohesiveness over simple graphs and the other is attributed community search (ACS) which concerns both the structural cohesiveness and content overlapping or similarities over attributed graphs.
Regarding capturing the structural cohesiveness, various community metrics have been proposed, including $k$-core~\cite{cs3,cs4,cs6}, $k$-truss~\cite{cs2,cs7}, $k$-clique \cite{cs1,cs8} and $k$-edge connected component~\cite{cs9,cs10}. 
These metrics are inflexible to adapt to complex real-world graphs and applications. 

In addition to only exploiting the structural information, ACS leverages both the structural constraint and attributes such as keywords~\cite{ACQ, ATC}, location~\cite{DBLP:conf/icde/WangCLZQ18}, temporal~\cite{DBLP:conf/icde/LiSQYD18}, etc. 
As two representative approaches for ACS, ATC~\cite{ATC} finds $k$-truss community with the maximum pre-defined attribute score. 
And ACQ~\cite{ACQ} finds $k$-core communities whose nodes share the maximum attributes with the query attributes.
Both ATC and ACQ adopt a two-stage process. First, they find the candidate communities based on the structural constraints.
Then, the candidates are verified based on the computed attribute score or the appearance of attribute set.
However, the quality of the found communities of the two approaches are unpromising since the independent two stages fail to capture the correlations between structures and attributes in a joint fashion. 

With the development of ML/DL, recently, GNN has been adopted for
CS~\cite{ICSGNN}.  By recasting the community membership determination
to a classification task, a model can learn via its prediction error
feedback given the training samples and can adapt to a specific graph
in an end-to-end way.  Recently, Gao et al. proposed ICS-GNN
~\cite{ICSGNN} for interactive CS, which allows users to provide
ground-truth for online incremental learning.  The model is a
query-specific model that fails to generalize to new query nodes.  
%
\cite{AQDGNN} proposes a graph neural network based model that is
trained by a collection of query nodes with their ground-truth, and
makes predictions for unseen query nodes in a single graph.

\stitle{ML/DL for Graph Analytics.} Apart from CS, ML/DL techniques are widely used in various graph analytical tasks, including classical combinatorial optimization problems~\cite{Combinatorial, Combinatorial2}, graph similarity search~\cite{TaGSim, GHashing, GLSearch}, subgraph matching~\cite{subiso, RLbased, NeuralSM}, subgraph counting~\cite{subgraphcounting, NSIC, zhao2023learned}, shortest path query~\cite{predict}, community collapsing~\cite{FindingCriticalUsers} and community detection~\cite{communitydetection}. 
In brief, the main ideas of these approaches contain learning a model-based algorithm heuristics~\cite{Combinatorial, Combinatorial2, GLSearch, RLbased} to replace the traditional predefined heuristics, where Reinforcement Learning algorithms can be used; learning a workload-specific estimator for approximate query processing~\cite{subgraphcounting, NSIC, TaGSim}; constructing a model-based database index for filtering or searching~\cite{subiso, NeuralSM, predict, GHashing, FindingCriticalUsers}, which is node or graph embedding preserving task-related semantics. 
Our approach is in the first category regarding learning a meta heuristic for CS while the subtle difference is that we leverage metric learning to evaluate the community membership.

\stitle{Meta-Learning on Graph.}
Meta-learning is a learning paradigm that learns the prior knowledge from multiple tasks, which can be swiftly transferred to a new task with only a few observed data. 
In general, the meta-learning approaches fall into three categories, i.e.,  black-box adaptation~\cite{NTM, meta1, MANN, meta4}, optimization-based~\cite{MAML, meta2, reptile}, and metric-based~\cite{prototypical, RN, MN} approaches. 
Meta-learning has been adopted over graph data to deal with various graph learning tasks, including node classification~\cite{metagnn, GMETA}, link prediction~\cite{metagraph, GMETA}, graph classification~\cite{5, DBLP:conf/cikm/MaBYZYYZY20} and graph alignment~\cite{62, 53}. 
A brief survey that summarizes the applications and methods can be found in~\cite{DBLP:journals/corr/abs-2103-00137}. 
Here, GNN is widely used as the base model or core component of these approaches. The optimization-based, metric-based, or hybrid of optimization and metric-based~\cite{GMETA} are used as the meta-learning strategies.
However, all the existing approaches are oriented to graph learning tasks and cannot be directly applied to our CS task, where the input is specified by a personalized query node.
Our approach CGNP can be regarded as metric-based since the predictive probability of CGNP is derived from inner-product similarity in a hidden embedding space. 

%% file: main.tex
\section{Problem Statement}
\label{sec:problem}

We consider an undirected simple graph ${\cal G} = (V, E)$, where
$V({\cal G})$ is the node set and $E({\cal G})$ is the edge set. Let
$n = |V({\cal G})|$ and $m = |E({\cal G})|$ denote the number of nodes
and edges, respectively.
The neighborhood of node $v$ is denoted as $\mathcal{N}(v) = \{ u |
(u, v) \in E({\cal G}) \}$.  The nodes may possess $d$ attributes
$\mathcal{A} = \{ \mathcal{A}_1, \cdots, \mathcal{A}_d\}$.  For each
node $v$, a one-hot $d$-dimensional vector $\mathcal{A}(v) \in \{0,
1\}^d$ encodes whether $v$ is associated with the $d$ attributes in
$\mathcal{A}$.
In the following, we use ${\bf G}$ to represent a large data graph.  A
\emph{community} in ${\bf G}$ is a cohesive subgraph $G = (V, E)$
induced by its node set $V(G)$, such that the nodes $V(G)$ are
intensively connected within $G$ whereas are sparsely connected with
other nodes in the graph, i.e., $|E(G)| \gg |\{ (u, v) | u \in V(G), v
\in V({\bf G}) \setminus V(G) \}|$.  Below, we denote a community as
an induced subgraph in a graph $G$ by $\community(G)$.

\stitle{Problem Statement (Community Search):}
The community search problem is to find the
query-dependent community $\community_q$, for a user-given query node
$q$ in a graph $G$, such that $q \in \community_q(G)$. Distinguished
from prior algorithmic approaches~\cite{ATC, ACQ, CTC}, the community
$\community_q(G)$ in this paper is not restricted in any $k$-related
subgraph, instead it is learned from given community membership
ground-truth.

We construct a meta model $\model$ to
support community search queries in a data graph ${\bf G}$ by multiple
tasks.  The model $\model$ is trained on a set of training tasks
$\mathcal{D} = \{ \task_i\}_{i = 1}^{N}$.  Here, a training task,
$\task_i$, is a triplet $\task = (G, Q, L)$, where $G$ is a subgraph
of ${\bf G}$, $Q= \{ q_1, \cdots, q_{j} | q_{i} \in V(G)\}$ is a set
of $j$ query nodes in $G$, and $L = \{ l_{q_1}, \cdots, l_{q_j}\}$ is
the ground-truth of the $j$ query nodes, respectively.  Specifically,
$l_{q}$ is a nonempty set of nodes in $G$ w.r.t. the query node $q$,
that contains a set of positive samples, $l_q^+ \subset
\mathcal{C}_{q}(G)$, and a set of negative samples, $l_q^{-} \subset
(V(G) \setminus \community_{q}(G))$.
 For a new test task $\task^*= (G^*, Q^*, L^*)$, the meta model $\model$ will exploit the query node set $Q^*$ associated
with the ground-truth $L^*$ to adapt to task $\task^*$, and
can make community search prediction for nodes in $V(G^*)\setminus Q^*$. Note that for test task, the number
of query nodes in $Q^*$, named shots, is rather limited, i.e., $|Q^*| \ll |V(G^*)|$.


It is important to mention that the main idea behind multiple tasks is
that it is difficult to obtain all required ground-truth to train.
There are many possible scenarios with different ways that the
training task set $\mathcal{D}$ and new test tasks are constructed.
%
%
In this paper, we construct tasks from two dimensions: Single/Multiple
graphs and Shared/Disjoint communities.
%
%
%


\begin{itemize}[noitemsep,topsep=0pt,parsep=5pt,partopsep=0pt,leftmargin=*]
\item {\sl Single Graph Shared Communities.}
The graphs in any training task, $G$, and test task, $G^*$, are
subgraphs of a single large graph ${\bf G}$. The query nodes in
training/test tasks are different but are from the same communities in
${\bf G}$.

\item {\sl Single Graph Disjoint Communities.}
The graphs in any training task, $G$, and test task $G^*$, are
subgraphs of a single large graph ${\bf G}$. The query nodes in
training/test task are from different communities in ${\bf G}$, such
that $\community_{q}(G) \cap \community_{q^*}(G^*) = \emptyset$, for
all $q \in Q$ and $q^* \in Q^*$.

\item {\sl Multiple Graphs Disjoint Communities.}
The graphs in any training task, $G$, and test task, $G^*$, are from
different large data graphs. The query nodes in training/test tasks
are from different communities.  Here, all the subgraphs $G$ in the
training tasks are from the same domain, whereas a subgraph $G^*$ in a
test task can be in the same or a different domain.
%
\end{itemize}

\begin{example}
Assume that the DBLP graph in Fig.~\ref{fig:case} is a single graph
${\bf G}$. A graph $G$ in a training task $\task = (G, Q, L)$ and a
graph $G^*$ in a test task $\task^* = (G^*, Q^*, L^*)$ are subgraphs
of ${\bf G}$.  Suppose a subgraph $G$ in a training task contains a
part of the community that `Jure' belongs to (i.e., the orange nodes)
in Fig.~\ref{fig:case}.  In the scenario of shared communities,
some nodes in $Q^*$ in a test task $\task^*$ may contain some
ground-truth (e.g., orange nodes) that do not appear in any training
tasks.  The model $\model$ trained is to find the community for any
query node in $Q^*$ in the same test task $\task^*$ without any
ground-truth associated with.
In the scenario of disjoint communities,
the ground-truth given in a test task may have nothing to do with the
community that 'Jure' belongs to.  The model $\model$ trained is to
find the community for any query node in $Q^*$ in the test task
$\task^*$ without any ground-truth associated with. Note that the
community to be found is a different community that 'Jure' belongs to.
For the scenario of multiple graphs, a subgraph $G$ in a training task
$\task$ is from a data graph, ${\bf G}$, whereas a subgraph $G^*$ in a
test task is from a different data graph ${\bf G}'$.
\comment{ Single Graph with
  Shared Communities, which indicates the training and test tasks have
  overlapped community information. If the test task does not contain
  any orange node of the community of `Jure', but contains partial
  yellow nodes in other communities, this is the scenario of
  Fig.~\ref{fig:task:sgdc}, Single Graph with Disjoint Communities. If
  $G$ and $G^*$ are different graphs, e.g., one is from DBLP graph and
  the other is ACM graph, this is the scenario of
  Fig.~\ref{fig:task:mgdc}, Multiple Graphs with Disjoint Communities.
}
\end{example}

\section{Naive Approaches}
\label{sec:naive}

\comment{
\begin{figure*}[t]
	\centering
	\begin{tabular}[h]{c}
		\subfigure[MAML: an Optimization-based View] {
			\includegraphics[ height=0.65\columnwidth]{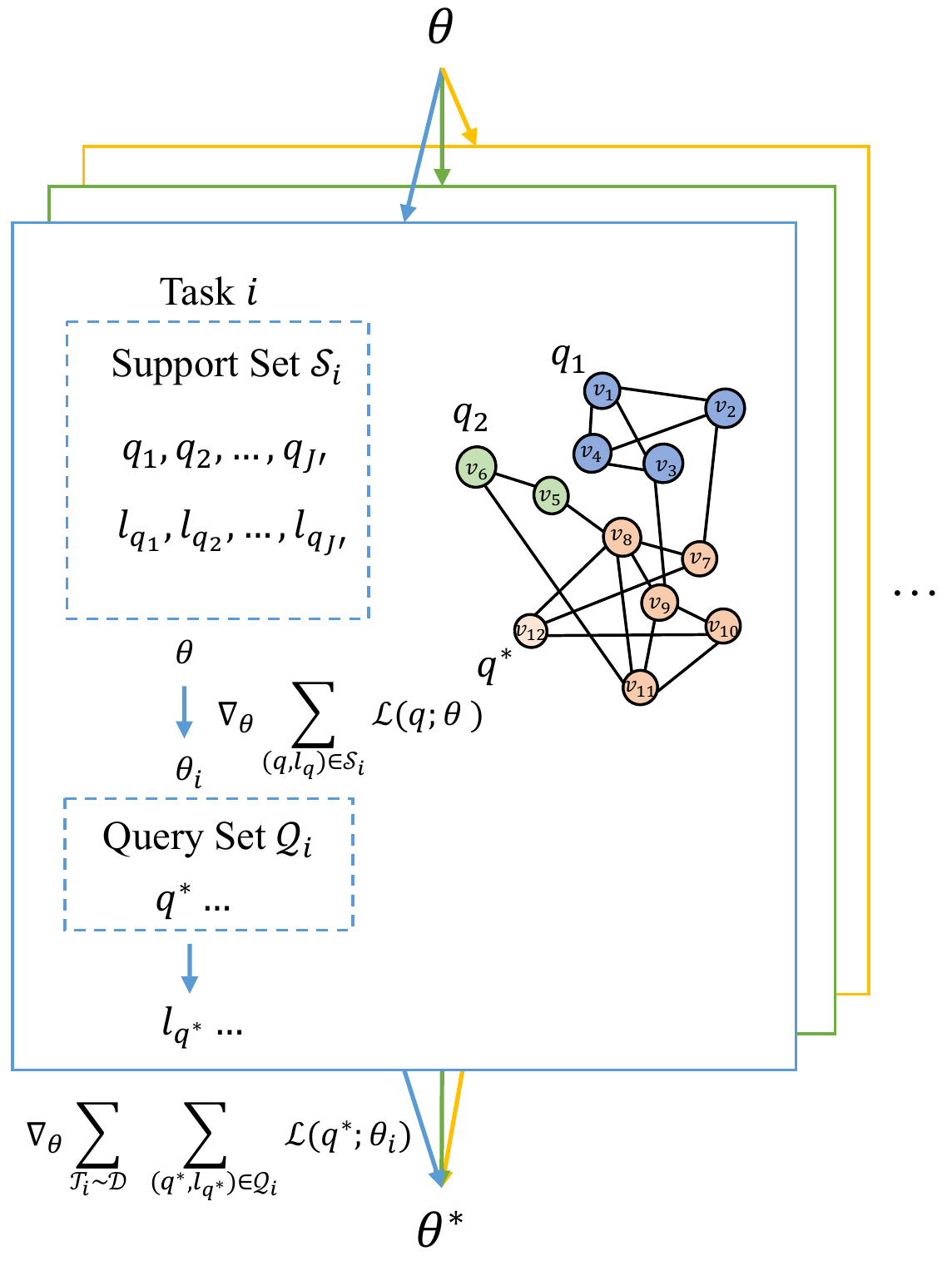}
			\label{fig:views:maml}
		}
		\hspace{-0.2cm}
		\subfigure[GPN: a Metric-based View] {
			\includegraphics[ height=0.65\columnwidth]{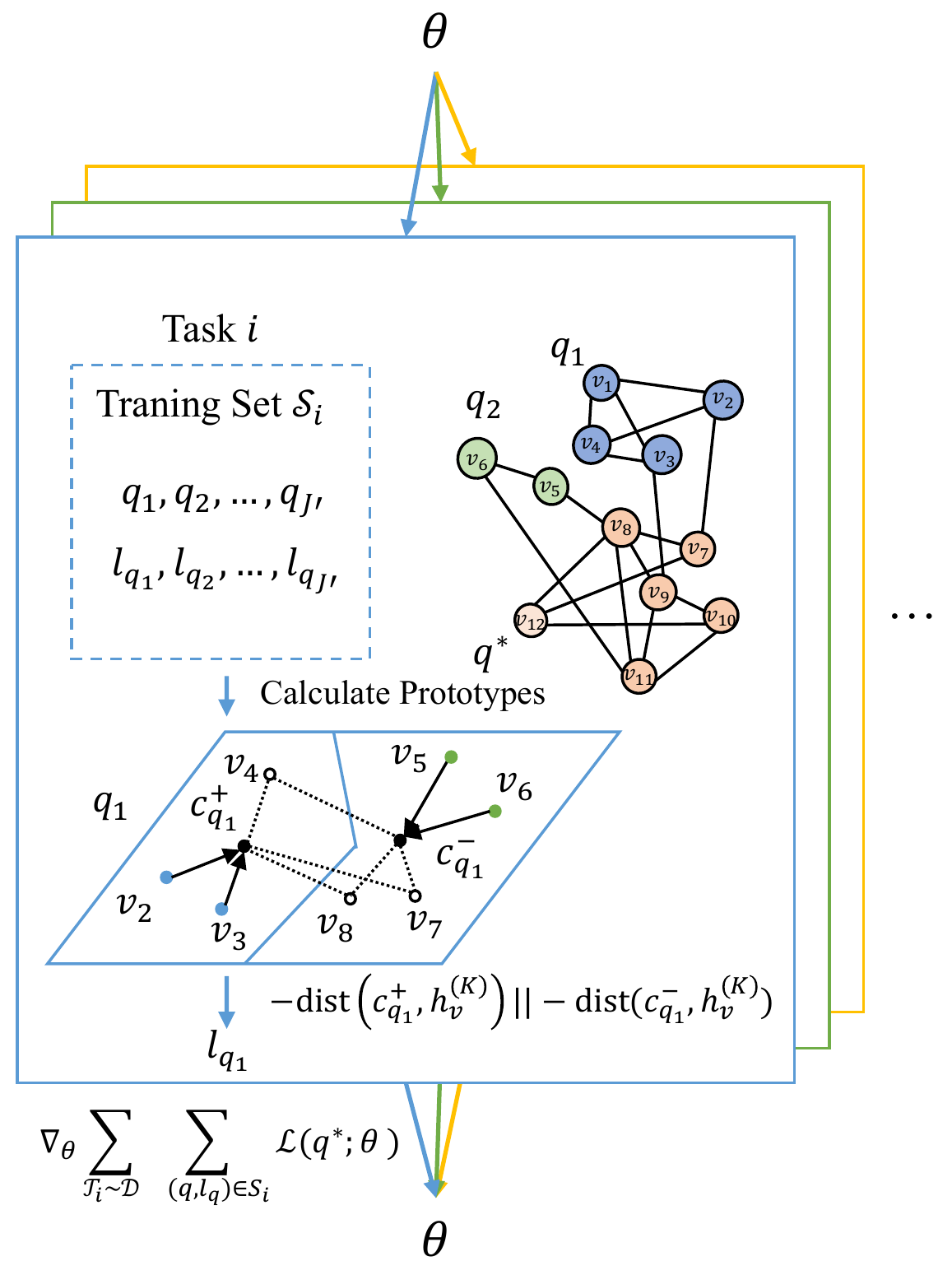}
			\label{fig:views:gpn}
		}
		\hspace{-0.2cm}
		\subfigure[CNP: a Kernel-based View] {
			\includegraphics[ height=0.65\columnwidth]{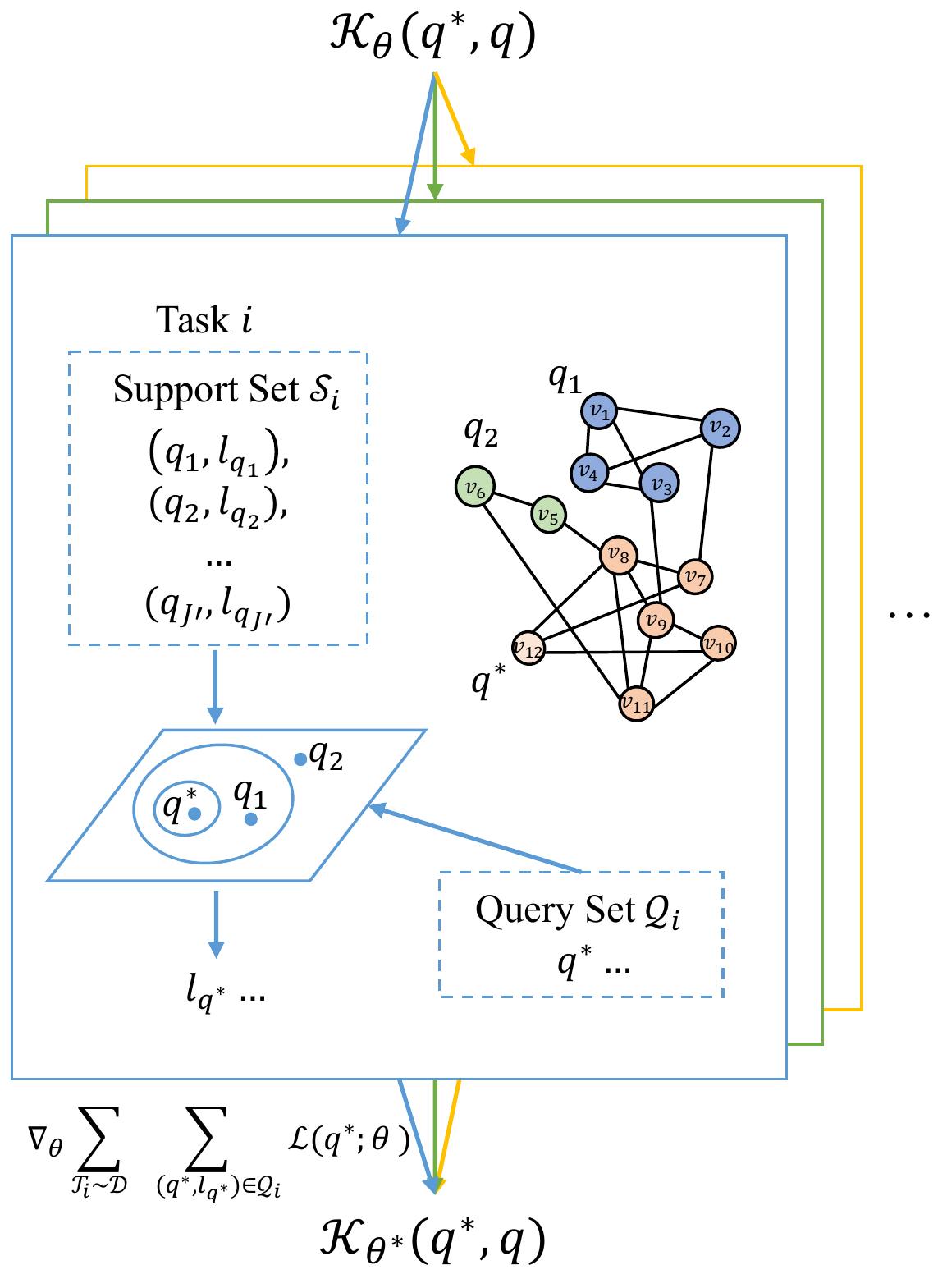}
			\label{fig:views:kernel}
		}
		\hspace{-0.2cm}
		\subfigure[CGNP: a Metric-based View] {
			\includegraphics[ height=0.65\columnwidth]{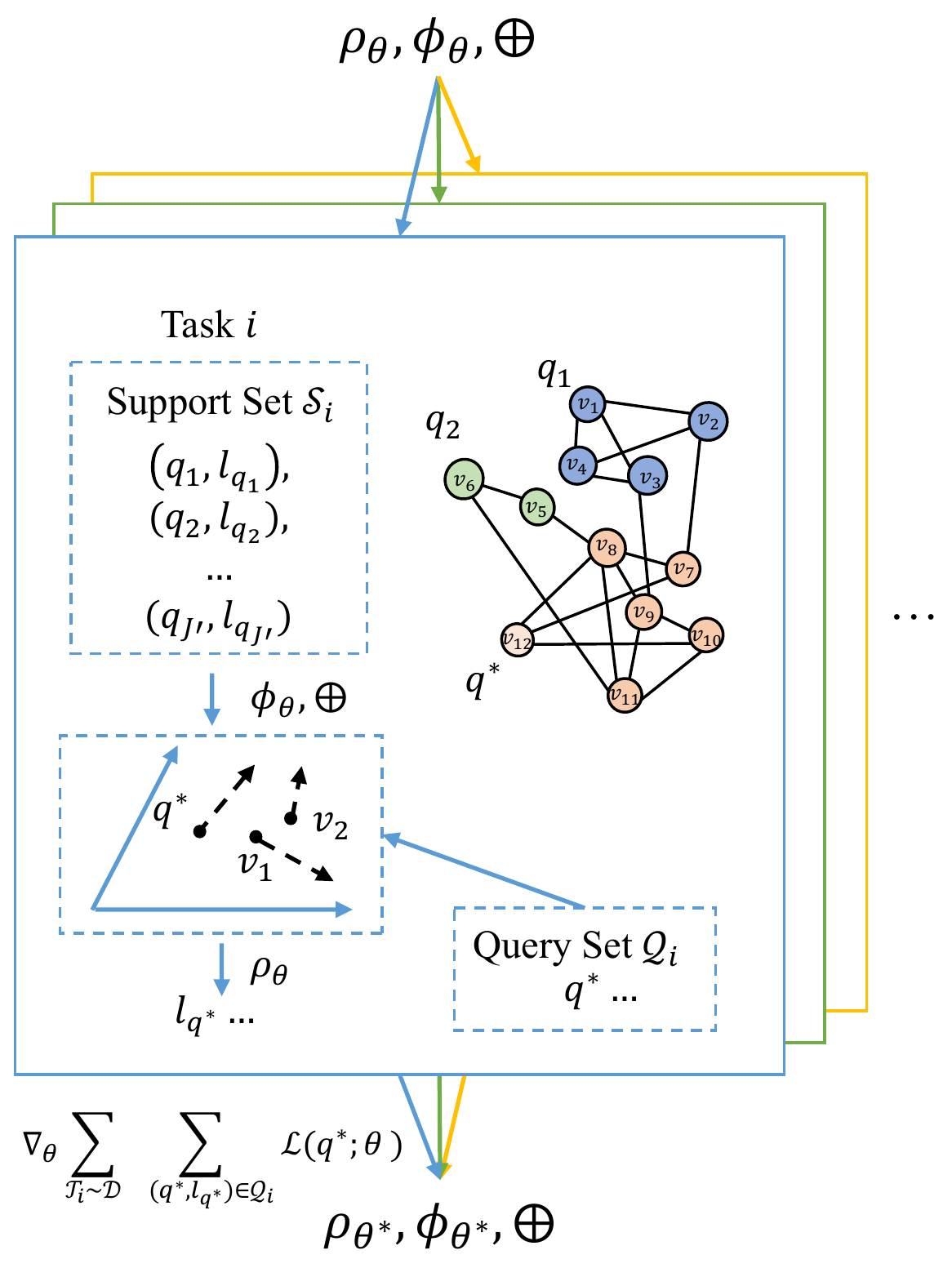}
			\label{fig:views:distance}
		}	
		
	\end{tabular}
	\vspace{-0.2cm}
	\caption{Intuitions of MAML, GPN, CNP and CGNP}
	\vspace{-0.2cm}
	\label{fig:views}
\end{figure*}
}

To construct a meta model, a naive approach is to pre-train a Graph
Neural Network (GNN) model over $\mathcal{D}$ and finetune the model
for a new task $\task^*$. Below, we first introduce multi-label
classification by GNN, which serves as the basis of the naive
approaches and our meta-learning approach.

Given a graph $G$, a $K$-layer GNN follows a neighborhood aggregation paradigm to generate a new representation for each node by aggregating the representations of its neighbors in $K$ iterations. 
Let $h_v^{(k)}$ denote the representation of a node $v$ generated in the $k$-th iteration, which is a $d^{(k)}$ dimensional vector. 
In the GNN $k$-th iteration (layer), for each node $v \in V(G)$, an aggregate function $\Fagg^{(k)}$ aggregates the representations of the neighbors of $v$ that are generated in the ($k$-$1$)-th iteration as Eq.~(\ref{eq:gnn:fagg}). 
Then, a combine function $\Fcom^{(k)}$ updates the representation of $v$ by the aggregated representation $a_v^{(k)}$ and previous representation $h_v^{(k-1)}$ as Eq.~(\ref{eq:gnn:fcom}). 
\begin{align}
	a_v^{(k)} &= \Fagg^{(k)}(\{ h_u^{(k - 1)} | u \in \neighbor(v)\}) \label{eq:gnn:fagg} \\
	h_v^{(k)}  &= \Fcom^{(k)}(h_v^{(k - 1)}, a_v^{(k)}) \label{eq:gnn:fcom}
\end{align}
The functions $\Fagg^{(k)}$ and $\Fcom^{(k)}$ are neural networks, e.g., linear transformation with non-linearities and optional Dropout for preventing overfitting. 
The neural network parameters from $\Fagg^{(k)}$ and $\Fcom^{(k)}$ are shared by all the nodes.

For a given task $\task = (G, Q, L)$, a GNN can be built by training
over $Q$ and $L$, then is deployed to make predictions for any query
node $q \in V(G) \setminus Q$ as a query.  Concretely, a binary query
identifier $I_q(v) \in \{ 0, 1\}$ is concatenated with the attribute
feature vector $\mathcal{A}(v)$ to form the initial node
representation $h_{v}^{(0)}$, where $I_q(v) = 1$ if $v$ is the query
node $q$ otherwise $I_q(v) = 0$. Through transformation of $K$ layers,
the 1-dimensional node representation $h_v^{(K)}$ is activated by a
$\sigmoid$~function, i.e., $\hat{y}(v) = \sigmoid(h_v^{(K)})$, which
is the likelihood that $v$ is in the same community with query node
$q$.  The given $Q$ and the ground-truth $L$ provide
the training data for the GNN model.  For a known node $q \in Q$ with
its ground-truth $l_q = (l_q^{+}, l_q^{-})$, where $l_q^{+}$ and
$l_q^{-}$ are the positive and negative samples respectively,
w.r.t. $q$, the binary cross entropy (BCE) loss in
Eq.~(\ref{eq:loss:bce}) evaluates the divergence between the
predictive probability of the nodes from the positive and negative
samples, under the GNN with parameter $\theta$.
\begin{align}
	\label{eq:loss:bce}
	\loss(q; \theta) = - \sum_{v^{+} \in l_q^+} \log\hat{y}(v^+) -  \sum_{v^{-} \in l_q^-} \log (1-\hat{y}(v^-)) 
\end{align}
Based on the simple GNN approach, we review three naive approaches
which are simple combinations of GNN and meta/transfer learning
algorithms. 

\stitle{Feature Transfer.}  The learned parameters of shallow layers in neural network 
can be transferred to new tasks, instead of learning from scratch. The intuition is that the pre-trained low-level feature transformation can be shared with a new task. 
Thereby, we can train a GNN by the union of all the $Q$ and $L$ of every training task $\task$ in the training set $\mathcal{D}$.
When a new task $\task^*$ arrives, the parameters of $\Fagg^{(K)}$ and $\Fcom^{(K)}$ will be updated by minimizing the BCE loss in Eq.~(\ref{eq:loss:bce}) over $Q^*$ and $L^*$ by several gradient steps. 
However, the effectiveness of simple feature transfer is limited. For one thing, this approach is originally proposed for convolutional neural network (CNN) to process image data, which has an explicit feature hierarchy to be transferred. However, whether the same transfer mechanism well suits GNN over graph data still needs exploration.  
For the other thing, it is hard to control the gradient steps in the fine-tuning procedure for various test tasks. 

\stitle{Model-Agnostic Meta-Learning (MAML).} A meta GNN model can be built by a model-agnostic meta-learning algorithm, MAML~\cite{MAML}, over a set of training tasks $\mathcal{D}$.
MAML is a two-level end-to-end optimization algorithm, where the lower level is to optimize task-specific parameters $\theta_i$ for one task $\task_i$ and the upper level is to optimize the task-common parameters $\theta^*$ over the training task set $\mathcal{D}$. 
The learned task-common parameters $\theta^*$ will be used as the neural network initialization and updated by a few gradient steps to generalize a new task $\task^*$, given the few-shot task-specific data $Q^*$ and $L^*$.
To be concrete, training data $Q_i = \{q_{j}\}_{j = 1}^{J}$ and $L_i = \{l_{q_j}\}_{j = 1}^{J}$ of one training task $\task_i$ are divided into two sets, $\support_i = \{(q_{j}, l_{q_j})\}_{j = 1}^{J'}$ and $\query_i = \{(q_{j}, l_{q_j})\}_{j = J' + 1}^{J}$. $\support_i$ is called \emph{support set} and $\query_i$ is called \emph{query set}. The task-specific parameters $\theta_i$ is updated by the support set of $\task_i$ as Eq.~(\ref{eq:maml:inner}) in an inner loop, and the task-common parameters $\theta^*$ is updated by the query set over $\mathcal{D}$ in an outer loop as Eq.~(\ref{eq:maml:outer}), by gradient descent with learning rates $\alpha$ and $\beta$, respectively. 
\begin{align}
	\theta_i & \leftarrow \theta - \alpha \nabla_{\theta} \sum_{(q, l_q) \in \support_i}\loss(q; \theta) \label{eq:maml:inner}\\
	\theta^*  & \leftarrow \theta -\beta \nabla_{\theta} \sum_{\task_i \sim \mathcal{D}} \sum_{(q, l_q) \in \query_i}\loss(q; \theta_i)  \label{eq:maml:outer}
\end{align}
Although MAML is an effective and fairly general framework, it suffers
from a variety of problems, including training instability,
restrictive model generalization performance and extensive
computational overhead~\cite{DBLP:conf/iclr/AntoniouES19}.  To
alleviate the computational overhead, Reptile, is proposed as a
first-order meta-learning algorithm~\cite{reptile}.  Reptile directly
updates the task-common parameters $\theta^*$ by the first-order
gradients, bypassing the computation of the high-order
derivatives. First, the inner loop follows MAML to compute the
task-specific parameters $\theta_{i}$ for $\task_i$ as
Eq.~(\ref{eq:maml:inner}). Then, in the outer loop, the task-common
parameters $\theta^*$ are directly updated by the difference of
$\theta_{i}$ to current parameters $\theta$ as shown in
Eq.~(\ref{eq:reptile:outer}).
\begin{align}
\theta^*  & \leftarrow \theta + \beta \frac{1}{|\mathcal{D}|}\sum_{\task_i \sim \mathcal{D}}(\theta_i-\theta) \label{eq:reptile:outer}
\end{align}
Here, since evaluating the query set $\query_i$ of $\task_i$ is unnecessary,
Reptile does not split $\query_i$ and $\support_i$ for updating $\theta_i$, but update $\theta_i$ by all the training data of $\task_i$ in the inner loop.

\stitle{Graph Prototypical Network (GPN).} Prototypical
Network~\cite{prototypical} is an effective approach for few-shot
classification, which learns a metric space in which classification
is performed by computing distances to the centroid (prototype)
representation of each class. Different from general classification,
the prototype representation of CS should be query-specific. For a
query node $q$, two prototypes, $c_q^{+}$ and $c_q^{-}$, are computed
by the mean representations of the positive and negative samples in
the ground-truth $l_q$, respectively (Eq.~(\ref{eq:gpn:proto})).
Here, $h_{v}^{(K)}$ is the node representation of $v$ of the $K$-th
layer of GNN, generated by Eq.~(\ref{eq:gnn:fcom}). Then, the
likelihood that node $v$ is in the same community with $q$ is
predicted by its distances to the prototypes as
Eq.~(\ref{eq:gpn:likelihood}), given a distance function \distance.
\comment{
\begin{align}
c_q^{+} = \frac{1}{|l_q^{+}|}\sum_{v^{+} \in l_q^{+}} f_{\theta}(v^{+}, G, q),~c_q^{-} = \frac{1}{|l_q^{-}|}\sum_{v^{-} \in l_q^{-}} f_{\theta}(v^{-}, G, q)  \\
\hat{y}(v) = \softmax \biggl( [-\distance(f_{\theta}(v, G, q), c_q^{+}) \| -\distance(f_{\theta}(v, G, q), c_q^{-})] \biggr)
\end {align}
}
\begin{align}
c_q^{+}  & = \frac{1}{|l_q^{+}|}\sum_{v^{+} \in l_q^{+}} h_{v^{+}}^{(K)},~c_q^{-} = \frac{1}{|l_q^{-}|}\sum_{v^{-} \in l_q^{-}} h_{v^{-}}^{(K)} \label{eq:gpn:proto} \\
\hat{y}(v) & = \softmax \biggl( [-\distance(h_{v}^{(K)}, c_q^{+}) \| -\distance(h_{v}^{(K)}, c_q^{-}) ] \biggr) \label{eq:gpn:likelihood}
\end {align}
In the training stage, ground-truth sets $l_q^{+}$ and $l_q^{-}$ are
split into two sets, respectively. One is used to compute the prototypes
in Eq.~(\ref{eq:gpn:proto}) and the other is to compute the BCE loss
in Eq.~(\ref{eq:loss:bce}) for parameter update.  It is worth
noting that each query node must compute its own prototypes as it has
its own communities.  Therefore, GPN cannot support query node in
the test task without any ground-truth, where computing prototypes is
infeasible.

\section{CGNP for CS}
\label{sec:metric}


To overcome the disadvantages of the naive approaches, 
we devise a novel meta-learning framework for CS, named Conditional Graph Neural Process (CGNP), on the basis of Conditional Neural Process (CNP)~\cite{CNP}.
In this section, we first introduce CNP as a preliminary, then present the core idea of CGNP for CS as an overview of our framework.

CNP is a neural network approximation of stochastic process, e.g., Gaussian Process (GP). It directly models the predictive distribution conditioned on an arbitrary number of context observations by neural networks.
Specifically, given observed data $X = \{x_i\}_{i = 1}^N$ with corresponding ground-truth $Y = \{ y_i\}_{i = 1}^N$, CNP models the predictive distribution of new data $x^*$ with the target $y^*$, $p(y^*| x^*, X, Y)$, by the neural network architecture in Eq.~(\ref{eq:cnp}).
\begin{align}
	\label{eq:cnp}
	p(y^*| x^*, X, Y) = \rho_\theta \biggl(x^*,  \bigoplus_{i = 1}^{N} \phi_\theta(x_i,y_i) \biggr) 
\end{align}
Here, $\phi_\theta: X \times Y \rightarrow \mathbb{R}^d$ and $\rho_\theta: X \times \mathbb{R}^d \rightarrow \mathbb{R}^e$ are neural networks. The big $\oplus$ is a commutative operation that takes elements in $\mathbb{R}^d$ and aggregates them into a single element of fixed length $\mathbb{R}^d$. 
$\phi_\theta$ is the encoder that transforms pairs of $(x_i, y_i)$ into $d$-dimensional hidden representations. 
The big $\oplus$ aggregates $N$ representations into a context representation in a permutation-invariant fashion which memorizes the whole dataset $X$ and $Y$. 
To deal with a query for new observation $x^*$,  a decoder $\rho_\theta$ takes the context and $x^*$ as inputs and makes a final prediction for $x^*$. 

Similar to stochastic process, CNP can be used to build meta models via learning the prior of data generation, where each data instance is a collection of $(x_i, y_i)$, i.e., a task. 
The difference lies in that stochastic process, e.g., GP, explicitly specifies the prior distribution, and optimizes the hyper-parameters of the prior by maximum likelihood. CNP instead explicitly parameterizes the predictive distribution as neural networks thereby learning the prior implicitly. 

%
\stitle{Conditional Graph Neural Process (CGNP).}
The CGNP model we propose is a graph specification of CNP for
query-dependent node classification. 
For a CS task $\task = (G, Q,
L)$, CGNP directly models the predictive probability $p(\hat{l_{q^*}}
| q^*, \task)$ for a new query node $q^* \in V(G) \setminus Q$, where
$\hat{l_{q^*}} = \{ \hat{l_{q^*}}(v) \}_{v \in V(G)} \in \{0, 1\}^{n}$
is the binary target prediction for all the nodes in $G$.  
We instantiate the encoder $\phi_{\theta}$ that encodes each query node $q$ with its ground $l_q$ in the task $\task$,
the commutative operation that generates the context representation of $\task$, and the encoder $\rho_{\theta}$ that predicts $p(\hat{l_{q^*}}| q^*, \task)$ as Eq.~(\ref{eq:cnp:kernel1}).
CGNP inherits the interpretation of CNP~\cite{CNP} that using neural networks to mimic an \emph{implicit kernel function}, $\Kernel(\cdot, \cdot)$, which evaluating the  similarity between an observed query node $q$ and the target query node $q^*$.
The predictive probability is the summation of the observed ground-truth $l_q$, weighted by the similarities of query nodes as Eq.~(\ref{eq:cnp:kernel2}).

\begin{figure*}[t] 
	\centering 
	\includegraphics[width=0.85\textwidth]{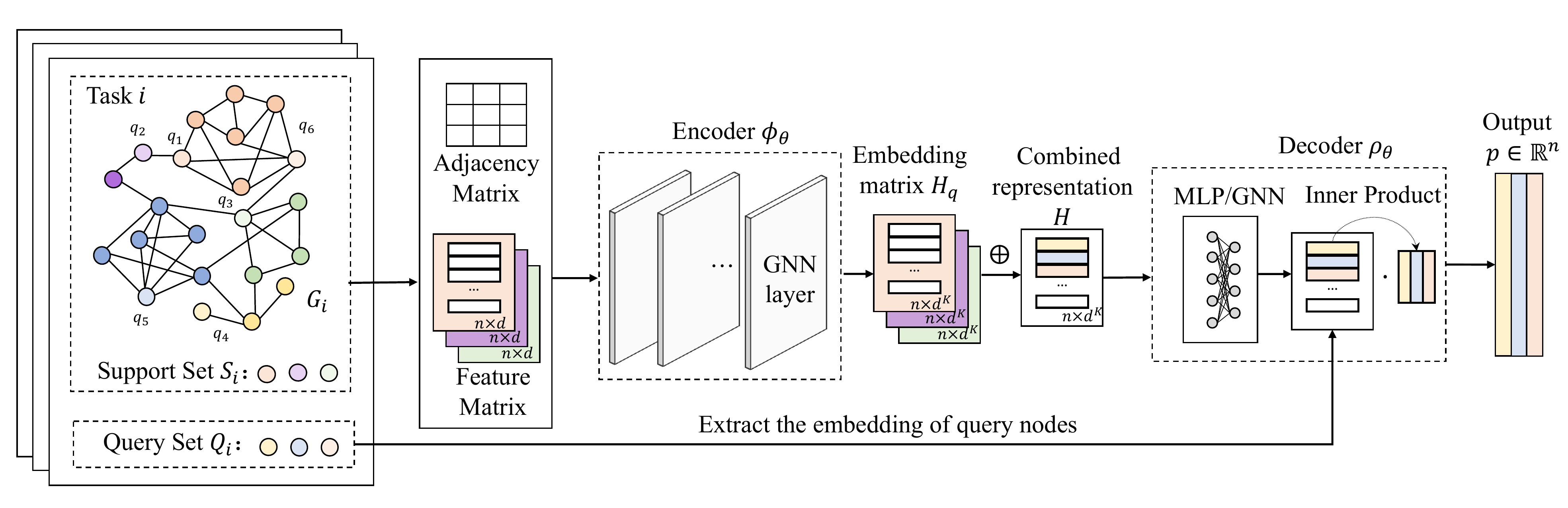} 
	\vspace{-0.3cm}
	\caption{The Architecture of CGNP} 
	\vspace{-0.5cm}
	\label{fig:cgnp} 
\end{figure*}

\begin{align}
\label{eq:cnp:kernel1}
p(\hat{l_{q^*}} |q^*, \task) &= \rho_\theta \biggl(q^*, \bigoplus_{(q, l_q) \in (Q, L)}  \phi_\theta(q, l_q) \biggr) \\
\label{eq:cnp:kernel2}
&\approx \sum_{(q, l_q) \in (Q, L)} \Kernel(q^*, q) \odot l_q 
\end{align}
Like k-nearest neighbor (KNN) algorithm, the non-parametric metric learning intuition makes CGNP promising for small samples and classification tasks as CS.
Unlike KNN, for CGNP, the kernel function $\Kernel(\cdot,\cdot)$, as well as the multiplication operation $\odot$ is implicitly learned from the data by our instantiated $\rho_\theta$, $\phi_\theta$ and big $\oplus$. 
In other words, KNN and CGNP memorize the input data in different ways. KNN persists the input data by simple concatenation whereas CGNP persists it by learning a hidden context representation.

Apart from the implicit kernel $\Kernel(\cdot, \cdot)$ derived from CNP, in particular, we also impose an explicit metric objective on the learning of CGNP. 
For CS task, since we only need a binary prediction to indicate whether a node $v$ is a community member of the query node $q^*$, we let the predictive probability of the membership be determined by the distance of $v$ and $q^*$ in a hidden space $H$ as Eq.~(\ref{eq:cnp:distance}). 
The mapping function from the initial node features to that hidden space is specified by the neural network encoder $\phi_\theta$, decoder $\rho_\theta$ and the commutative operation big $\oplus$ of CNP.
%
\begin{align}
\label{eq:cnp:distance}
p(\hat{l_{q^*}}(v) | q^*, \task) &\models \distance( H(q^*), H(v)) 
\end{align}
In this explicit metric-based modeling perspective, we use inner product similarity to distinguish between membership and nonmembership in the hidden space $H$ that is learned in the training process.
The meta CGNP model is also a common neural network mapping, shared by multiple CS tasks, that maps nodes for partitioning in a task-specific hidden space $H$.
Note the kernel $\Kernel(\cdot,\cdot)$ in Eq.~(\ref{eq:cnp:kernel2}) and the distance $\distance(\cdot, \cdot)$ in Eq.~(\ref{eq:cnp:distance}) are two different concepts. 
The kernel measures the similarity between two input query nodes $q$ and $q^*$ regarding their community-member relationship with all the remaining nodes in $G$, whereas the distance measures the closeness of a query node $q$ and one remaining node $v$ in $G$ regarding their community membership.  
To learn a task-common kernel and distance mapping, CGNP iterates on the training task set to optimize the neural network parameters $\theta$, where data in one task is processed as a batch.
Compared with the two-level optimization-based algorithm MAML, 
CGNP learns the prior knowledge of CS by metric learning for node clustering/partitioning, which better exploits small data for classification and avoids unstable and inefficient parameter adaptation in the test stage.
The metric learning principle of CGNP is also different from that of GPN. 
GPN computes positive and negative prototypes, $c_q^{+}$ and $c_q^{-}$, for each query node $q$ by its ground-truth. Inference on other nodes is based on their distances to the prototypes. 
In contrast, CGNP directly models and evaluates the distances between the query nodes and the remaining nodes, thereby supporting queries without any ground-truth in test tasks.

\comment{
\textcolor{blue}{
CGNP is a metric-based learning approach that mimics a non-parametric model, Gaussian Process. Predictions are made by evaluating the distances between the embedding of the query nodes and other nodes. This metric-based learning strategy is more robust to the problem of imbalanced labels, achieving a higher recall than direct binary classification. That's the main reason that CGNP can outperform other naive approaches for CS.
}
}

\section{CGNP Model Design}
\label{sec:CGNP}

\comment{
A meta CGNP model $\mathcal{M}$ over multiple tasks should achieves the minimize conditional log probability over the probability of tasks $\mathcal{D}$. 
\begin{align}
\mathcal{M} &= \arg \min_{\theta} -\mathbb{E}_{\task \sim \mathcal{D}} [ \mathbb{E}_{q^*} [\log p(\hat{y}| q^*, \task)] ] 
\end{align}
}
We elaborate on how to design a CGNP model for the query-dependent node classification over graphs.
CGNP adopts an encoder-decoder based architecture and operates on task-level. Fig.~\ref{fig:cgnp} delineates the architecture of CGNP, which is composed of a GNN based encoder operating on query-level representation, a commutative operation, big $\oplus$, combining query-level representation to task-level context, and a decoder to perform final predictions.

\stitle{GNN Encoder ($\phi_{\theta}(q, l_q, G)$).} For each query node $q \in Q$ and its corresponding ground-truth $l_q \in L$, the encoder $\phi_{\theta}(q, l_q, G)$ is a $K$-layer GNN that maps the pair $(q, l_q)$ together with the graph $G$ to a node embedding matrix $H_q = \{ h_{v}^{(K)}\}_{v \in V(G)} \in \mathbb{R}^{n \times d^K}$. 
Here, $h_{v}^{(K)}$ is a $d^K$-dimensional output of the $K$-th layer of GNN for node $v$. 
The subscript $q$ of $H_q$ indicates the node embedding $H_q$ is generated particularly for query node $q$, as all the query nodes in $Q$ share the same GNN encoder.
Specifically, as the inputs of GNN, the adjacency matrix of graph $G$ is used for message passing of GNN, and $(q, l)$ determines the initial node $h_v^{(0)}$ as Eq.~(\ref{eq:cnp:encoder:feats}), where $\|$ is the vector/bit concatenation operation, $\mathcal{A}(v)$ is the attribute features of node $v$. In Eq.~(\ref{eq:cnp:encoder:feats}), $I_l(v) \in \{0, 1\}$ is a binary ground-truth identifier which distinguishes nodes within and without a same community, under the close world assumption. 
\begin{align}
	\label{eq:cnp:encoder:feats}
	h_v^{(0)} &= [ I_l(v) \| \mathcal{A}(v) ],
	I_l(v) =
	\begin{cases}
		1 & {v \in l_q^{+} \cup \{q\}  }  \\
		0 & \text{otherwise}
	\end{cases}
\end{align}
We can concatenate auxiliary features, e.g., the core number and local clustering coefficient of node $v$ on $h_v^{(0)}$ to exploit additional structural information.
The intuition of the GNN encoder is to generate a view, $H_q$, for the whole graph given an observation $(q, l_q)$ by message passing. 
A collection of views will be aggregated by the commutative operation big $\oplus$. This idea is enlightened by a CNP specialization for 3D scene understanding and rendering, Generative Query Network (GQN)~\cite{GQN}, where few-shot observed 3D views are summed up for predicting the view of a new query perspective. It is worth mentioning that, to the best of our knowledge, we are the first to introduce the insight of GQN to graph domain. 

\stitle{Commutative Operation ($\oplus$).} To combine the views $H_q$ for all query nodes in $ Q$ into one context representation, CGNP is equipped with three choices of commutative operations, sum, average and self-attention. All of the three operations are permutation-invariant. 

\etitle{Sum \& Average} are simple yet widely used pooling operations in many CNP instances~\cite{GQN, CNP}. The sum operation conducts element-wise sum up as Eq.~(\ref{eq:cnp:pool:sum}) and average further imposes a denominator of $|Q|$.
%
\begin{equation}
	\label{eq:cnp:pool:sum}
	H = \sum_{q \in Q} H_q
\end{equation}

\etitle{Self-Attention} is inspired by Attentive Neural Process (ANP)~\cite{DBLP:conf/iclr/KimMSGERVT19} and GP. 
Instead of giving the same weight to aggregate multiple data points, ANP and GP aggregate observed data by self-adaptive weights by self-attention~\cite{attention} and GP's kernel function, respectively.  
Thereby, CGNP leverages the self-attention to combine the node representations derived from all the query nodes, weighted by a set of learnable weights $\{w_q \}_{q \in Q} \in \mathbb{R}^{|Q|}$ as $H = \sum_{q \in Q} w_q H_q$. The weights $\{w_q \}_{q \in Q}$ are shared by all the nodes in $G$. 
\comment{
\begin{align}
	\label{eq:cnp:pool:attention}
	
\end{align}
}
Specifically, to compute the attention weight $\{w_q \}_{q \in Q}$ by the multiple views $\{ H_q\}_{q \in Q}$, let $\mathcal{H} = \{H_q[v]\}_{q \in Q} \in \mathbb{R}^{|Q| \times d^K}$ be the matrix stacked by the $|Q|$ node embeddings in $\{ H_q\}_{q \in Q}$ for an arbitrary node $v$. 
In Eq.~(\ref{eq:cnp:pool:att:trans}), 
$\mathcal{H}_{1} , \mathcal{H}_{2} \in \mathbb{R}^{|Q| \times d'}$ are transformed by linear weight matrices $W_{1} , W_{2} \in \mathbb{R}^{d^{K} \times d'}$, respectively. 
$\{w_q \}_{q \in Q}$ is computed by the inner product of $\mathcal{H}_{1}$ and the transpose of $\mathcal{H}_{2}$ followed by a $\softmax$ function that normalizes the weights to a probability, as Eq.~(\ref{eq:cnp:pool:att:weight}) shows.
\begin{align}
	\label{eq:cnp:pool:att:trans}
	\mathcal{H}_{1} &= \mathcal{H} W_{1},~ \mathcal{H}_{2} = \mathcal{H} W_{2},~\\
	\label{eq:cnp:pool:att:weight}
	\{w_q \}_{q \in Q} &= \softmax \biggl(\frac{ \langle\mathcal{H}_{1},  \mathcal{H}_{2}^T \rangle}{\sqrt{d'}} \biggr) 
\end{align}

\stitle{Decoder ($\rho_{\theta}(q^*, H)$).} Given the combined context $H$, a decoder $\rho_{\theta}(q^*, H)$ estimates the membership  for a new query node $q^*$, $p(l^* | q^*, H) \in \mathbb{R}^{n}$, conditioned on the memorized context $H$. 
We design three decoders with different complexities, a simple inner product decoder, multi-layer perception (MLP) decoder and GNN decoder. The latter two decoders MLP and GNN are also based on inner product. 

\etitle{Inner Product Decoder} is free of parameters and only operates on the context $H$. Since $H$ is a node embedding combined by multiple views, we can directly compute the node similarities between the embedding of a query node $q$ and all the other nodes. We use the inner product operation, $\langle \cdot, \cdot \rangle$, to compute the similarity score as Eq.~(\ref{eq:cnp:decoder:product}), followed by a $\sigmoid$ function to predict the probability that one node is in the same community with query node $q^*$.
The inner product operation indicates that the smaller the angle of two node embeddings in the vector space, the more likely the two nodes are from the same community.  
\begin{align}
	\label{eq:cnp:decoder:product}
	p(\hat{l_{q^*}} | q^*, \task) = \sigmoid(\langle H[q^*], H \rangle) 
\end{align}

\comment{
\etitle{MLP Decoder} firstly transforms the context matrix $H$ by a two-layer MLP as  Eq.~(\ref{eq:cnp:decoder:mlp}), then feeds the transformed $H$ to an inner product decoder of Eq.~(\ref{eq:cnp:decoder:product}). 

\begin{align}
	\label{eq:cnp:decoder:mlp}
	H \leftarrow \relu(H W_1) W_2
\end{align}

\etitle{GNN Decoder} firstly transforms the context matrix $H$ by a $K$-layer GNN as Eq.~(\ref{eq:cnp:decoder:gnn}), then feeds the transformed $H$ to an inner product decoder of Eq.~(\ref{eq:cnp:decoder:product}). In Eq.~(\ref{eq:cnp:decoder:gnn}), we use $W = \{ \Fagg^{(0)}, \Fcom^{(0)}, \cdots, \Fagg^{(K)}, \Fcom^{(K)} \}$ to denote the weights in the $K$-layer GNN. Note the GNN here is independent to the GNN in the encoder. 
\begin{align}
	\label{eq:cnp:decoder:gnn}
	H \leftarrow \kw{GNN}(H, G, W)
\end{align}
}

\etitle{MLP \& GNN Decoder.} MLP decoder firstly transforms the context matrix $H$ by an MLP, then feeds the transformed $H$ to an inner product operation of Eq.~(\ref{eq:cnp:decoder:product}). Similarly, 
firstly transforms the context matrix $H$ by a $K$-layer GNN, followed by the inner product. Note the GNN here is independent to the GNN in the encoder.
In contrast to the inner product decoder, the MLP and GNN encoder impose additional parametric transformations on the combined context embedding $H$ to improve the modeling capability of the decoder. The difference between MLP and GNN lies in that GNN further allows message passing among the nodes whereas the MLP transforms each node independently. 

\comment{
\textcolor{blue}{
	\begin{example}
		Take Single Graph with Shared Communities (SGSC) for example, given the Test Task \{12-Deepayan, 16-Julian, 32-Jaewon, 92-PaeaLe, 100-Jeff\} with query node $16-Julian$, the GNN Encoder maps the pair $(q, l_q)$ to the node embedding $H_q \in \mathbb{R}^{n \times d^K}$, where $q=16$ and $n=5$. Then combine the views of several different query nodes $H_q$ and get combined context $H$. Then It will transform by different decoder type. A new query node $(q, l_q)$ where $q=32$ is given, use inner product to measure the similarity score, that is $\langle H[q], H\rangle$. Followed by a sigmoid function, we predict the probability of node $\{12, 16, 92, 100\}$ is in the same community with query node $q=32$.
	\end{example}
 }
 }
 
In the following, we present the learning algorithms to train a meta CGNP model $\model$ and adapt the model to new tasks.
Recall that CGNP is to model a generative process of tasks $f \sim \mathcal{D}$, where $\mathcal{D}$ is the set of training tasks $\{ \task_i \}_{i = 1}^{N}$. Suppose the tasks are independent and the query nodes are independent in each task. The marginal likelihood of CGNP over $\mathcal{D}$ is 
\begin{align}
	p(\{L_1, \cdots, L_N\} | \{Q_1, \cdots, Q_N\}, \theta) = \prod_{\task_i \in \mathcal{D}} p(L_i | Q_i) 
\end{align}
Similar to MAML, for one training task $\task_i$, we split the training data $Q_i = \{ q_{j}\}_{j = 1}^{J}$ and $L_i = \{ l_{q_j} \}_{j = 1}^{J}$ into the support set $\support_i = \{ (q_{j}, l_{q_j})\}_{j = 1}^{J'}$ and query set $\query_i = \{ (q_{j}, l_{q_j})\}_{j = J' + 1}^{J}$.
The learning objective is to minimize the negative log-likelihood of the query set $\query_i$ conditioned on the support set $\support_i$ across all the tasks in $\mathcal{D}$ as Eq.~(\ref{eq:loss:nll}). 
The negative  log-likelihood loss in Eq.~(\ref{eq:loss:nll}) is in accordance with the BCE loss (Eq.~(\ref{eq:loss:bce})) of the query nodes in the query set $\query_i$. 
\begin{align}
	\label{eq:loss:nll}
	\loss &= -\sum_{\task_i \in \mathcal{D}} \sum _{(q, l_q) \in \query_i} \log p(l_q | q, \support_i) \\
	\nonumber
	&= -\sum_{\task_i \in \mathcal{D}} \!\sum _{(q, l_q) \in \query_i} \biggl (\sum_{v^{+} \in l_q^+} \log\hat{y}(v^+) +  \sum_{v^{-} \in l_q^-} \log (1-\hat{y}(v^-)) \biggr ) 
\end{align}

\begin{algorithm}[t]
	\footnotesize
	\caption{CGNP Meta Train}
	\label{alg:train}
	\DontPrintSemicolon
	\SetKwData{Up}{up}  \SetKwInOut{Input}{Input} \SetKwInOut{Output}{Output}
	\Input{training task set $\mathcal{D} =\{\task_i\}_{i = 1}^{N}$, learning rate $\alpha$, number of epochs $T$ }
	\Output{parameters $\theta$ of meta model $\mathcal{M}$}
	\SetKwFunction{Emit}{Emit}
	\SetKwFunction{Check}{Check}
	
	\For{$epoch \leftarrow 1$ to $T$}{ \label{line:epoch:start}
		Shuffle the task set $\mathcal{D} =\{\task_i\}_{i = 1}^{N}$; \label{line:shuffle} \; 	
		\For { $\task_i=(G_i, Q_i, L_i) \in \mathcal{D}$ } 
		{ 
			$\support_i, \query_i \sim (Q_i, L_i)$;  \label{line:support-query:split}   {\shadd{\tiny $\rhd$ allocate support and query sets} }\;
			\For {$(q, l_q) \in \support_i$}
			{ 
				\label{line:train:encoder:start}
				$H_q\leftarrow \phi_\theta(q, l_q, G_i)$; \label{line:train:encoder:end}  {\shadd{\tiny $\rhd$  compute query-specific view}} \;
			}
			$H \leftarrow \bigoplus_{(q, l_q) \in \support_i} H_q$; \label{line:train:agg} {\shadd{\tiny $\rhd$ compute context embedding (Eq.~(\ref{eq:cnp:pool:sum}))}}  \;
			\For {\ $(q, l_q) \in \query_i$}
			{
				$p(\hat{l_q}|q, \support_i) \leftarrow \rho_{\theta}(q, H)$; \label{line:train:decoder} {\shadd{\tiny $\rhd$ compute pred. prob. (Eq.~(\ref{eq:cnp:decoder:product}))}} \;
				Compute the Loss $\mathcal{L}(q)$ by $p(\hat{l_q}|q, \support_i)$ and $l_q$; \label{line:train:loss} \;
			}
			$\mathcal{L} \leftarrow \sum_{(q, l_q) \in \query_i}\mathcal{L}(q)$; \label{line:train:taskloss} \;
			
			$\theta \leftarrow \theta-\alpha \nabla_{\theta} \mathcal{L}$; \label{line:epoch:end}  {\shadd{\tiny$\rhd$  update model parameters}}\; 
	}}
	\Return{$\theta$};\
\end{algorithm}

\begin{algorithm}[!t]
	\footnotesize
	\caption{CGNP Meta Test}
	\label{alg:test}
	\DontPrintSemicolon
	\SetKwData{Up}{up}  \SetKwInOut{Input}{Input} \SetKwInOut{Output}{Output}
	\Input{test task $\mathcal{T^*}=(G^*, Q^*, L^*)$, parameter $\theta$ of meta model $\mathcal{M}$, a query node $q^* \in V(G^*) \setminus Q^*$}
	\Output{predictive probability of $q^*$}
	\SetKwFunction{Emit}{Emit}
	\SetKwFunction{Check}{Check}
	$\support^* \leftarrow (Q^*, L^*)$; \label{line:test:support-query:split} 
	
	\For { $(q, l_q) \in \support^*$}
	{ \label{line:test:encoder:start}
		$H_q\leftarrow \phi_\theta(q,l_q, G^*)$; \label{line:test:encoder:end}  {\shadd{\tiny $\rhd$  compute query-specific view}} \; 
	}
	$H \leftarrow \bigoplus_{q \in \support^*} H_q$;\label{line:test:agg}{\shadd {\tiny $\rhd$ compute context embedding (Eq.~(\ref{eq:cnp:pool:sum}))}} \;
	$p(\hat{l_{q^*}} | q^*, \support^*) \leftarrow \rho_\theta(q^*, H)$; \label{line:test:decoder}  {\shadd{\tiny $\rhd$ compute pred. prob. (Eq.~(\ref{eq:cnp:decoder:product}))} }\; 
	\Return{$p(\hat{l_{q^*}} | q^*, \support^*)$}; \;
\end{algorithm}

\stitle{Meta Training.} 
In the training stage, given the training task set $\mathcal{D}$, learning rate $\alpha$, and the number of epochs $T$, a meta CGNP model is trained by optimizing the negative log-likelihood of Eq.~(\ref{eq:loss:nll}) by stochastic gradient descent. 
Algorithm~\ref{alg:train} presents the training process. 
In each epoch (line~\ref{line:epoch:start}-\ref{line:epoch:end}), all the training tasks are randomly shuffled in line~\ref{line:shuffle}.
For each task $\task_i$, we get the allocated support set $\support_i$ and query set $\query_i$ from the given query node and ground-truth (line~\ref{line:support-query:split}).  
First, each query node $q$ associated with the ground-truth $l$ in the support set $\support_i$, together with the graph structure $G_i$ is fed into the GNN encoder, $\phi_\theta$ , to generate a query-specific view $H_q$ (line~\ref{line:train:encoder:start}-\ref{line:train:encoder:end}).
Second, in line~\ref{line:train:agg}, all the views are aggregated into the context matrix $H$ by the permutation-invariant operation big $\oplus$, e.g., by the summation aggregation of Eq.~(\ref{eq:cnp:pool:sum}). 
Third, for each query node in the query set $\query_i$, 
we compute its predictive probability and loss in line~\ref{line:train:decoder}-\ref{line:train:loss}, via evaluating the inner product similarities between node presentations and the query representation in $H$ as Eq.~(\ref{eq:cnp:decoder:product}).
Fourth, the model is updated by one gradient step of the aggregated task-specific loss (line~\ref{line:train:taskloss}-\ref{line:epoch:end}).

\stitle{Meta Testing.} For a test task $\task^*$ with graph $G^*$, few-shot query nodes $Q^*$ and the associated ground-truth $L^*$, Algorithm~\ref{alg:test} presents the steps to predict the community members for a query node $q^*$. The whole $Q^*$ and $L^*$ serve as the support set $\support^*$ (line~\ref{line:test:support-query:split}), followed by computing the context representation $H$ (line~\ref{line:test:encoder:start}-\ref{line:test:agg}). 
Finally, the query node $q^*$ and context $H$ are fed into the decoder network $\rho_\theta$ to obtain the prediction.

\begin{example}
We use a real example to illustrate how CGNP works in
Algorithm~\ref{alg:test} on a test task $\task^*= (G^*, Q^*, L^*)$ of
a DBLP subgraph, $G^*$, in Fig.~\ref{fig:case}. Suppose the task
possesses query nodes $Q^* =\{q_1, q_2, q_3\}$ correspond to 3 users
\{Julian, Jaewon, Deepayan\}, respectively, and the corresponding
ground-truth $L^* =\{ l_1, l_2, l_3\}$.  Each $l_i$ is composed a
positive node set $l_i^+$, and a negative node set $l_i^-$.  First,
for the 3 pairs $(q_1, l_1), (q_2, l_2), (q_3, l_3)$, by
line~\ref{line:test:encoder:start}-\ref{line:test:encoder:end}, the
GNN encoder $\phi_{\theta}$ generates 3 node embedding matrices $H_1,
H_2, H_3 \in \mathbb{R}^{n \times d^K}$ as the query-specific views,
respectively, where $n = |V(G)|$.  Second, by
line~\ref{line:test:agg}, the combine operator big $\oplus$ aggregates
the three matrices $H_1, H_2, H_3$ to one context embedding $H$.
Given a query node $q^*$ in the subgraph $G$, e.g, Jure, by
line~\ref{line:test:decoder} the inner product decoder $\rho_{\theta}$
predicts the probability of community membership of Jure for all the
nodes, by computing the inner product similarity of vector $H[q^*]$
and $H$ followed by a \kw{sigmoid} function as
Eq.~(\ref{eq:cnp:decoder:product}).
\end{example}

\stitle{Computation Complexity.}
 We analyze the time complexity of CGNP in brief. To be concise, we assume fixed dimension vector add, multiplication, and inner product take constant time when the dimension is far smaller than the graph node number $n$. 
For the GNN encoder of CGNP, the time complexity is $\mathcal{O}(K m |\support|)$ for a single task, where $K$ is the number of GNN layers, $m$ is the number of edges and $|\support|$ denotes the number of shots.   
The complexity of the big $\oplus$ operation is $\mathcal{O}(n |\support|)$ for the sum and average pooling and $\mathcal{O}(n |\support|^2)$ for the self-attention, respectively.  
For the decoders, the inner product operation takes $\mathcal{O}(n |\query|)$ time, and an MLP decoder and $K'$ layer GNN decoder takes extra $\mathcal{O}(n |\query|)$ and $\mathcal{O}(K' m |\query|)$ cost, respectively.
In total, the complexity of the meta test algorithm, Algorithm~\ref{alg:test},  is $\mathcal{O}(c (n + m))$, where $c$ is a constant determined by $K, K', |\support|, |\query|$. And the training complexity of Algorithm~\ref{alg:train} is $\mathcal{O}(TNc(n + m))$, where $T$ and $N$ are the numbers of iterations and training tasks. 

%% file: experiment.tex
\section{Experimental Studies}
\label{sec:exp}

We introduce the experimental setup (\cref{sec:exp:setup}) and report our  substantial results as follows:
\ding{172} compare the effectiveness of CGNP under different task configurations (\cref{sec:exp:effect}),
\ding{173} evaluate the efficiency of CGNP with the baselines, and conduct scalability test for learning-based approaches (\cref{sec:exp:effic}),
\ding{174} investigate the effect of the volume of the ground-truth on the performance of CGNP (\cref{sec:exp:labels}), and 
\ding{175} conduct the ablation studies on the CGNP model regarding the GNN layer and the commutative operation (\cref{sec:exp:ablation}).

\subsection{Experimental Setup}
\label{sec:exp:setup}

\stitle{Datasets:} We use 6 real-world graph datasets, including five
single graphs (\Cora, \Citeseer, \Arxiv, \Reddit, \DBLP) and one
multiple graph (\Facebook).  Table~\ref{tab:dataset} lists the profile
of the 6 datasets.
\Cora, \Citeseer and \Arxiv are citation networks whose nodes
represent research papers and edges represent citation
relationships. We use node class labels to simulate the communities
derived from the paper citation, which reveal the research topics that
papers belong to.
\DBLP \cite{DBLP} is a co-authorship network where nodes represent
authors and two authors are connected if they collaborate on at least
one paper. A ground-truth community is by the publication venue.
%
%
\Reddit is collected from an online discussion forum, where nodes
refer to posts, and an edge between two posts exists if a user
comments on both of the posts.  The ground-truth is the communities
that posts belong to.
\Facebook is a dataset containing 10
ego-centric social networks, which have friendship community
ground-truth.
\Cora, \Citeseer, and \Facebook have discrete node attributes. The
attributes of \Cora and \Citeseer are the keywords in the papers and
the attributes of \Facebook are the user properties.  For \Cora,
\Citeseer, and \Facebook, we use one-hot representations of the
attributes as the node features, concatenating with the core number
and local cluster coefficient of the node.  We use core number and
local cluster coefficient alone as node features, for \Arxiv, \DBLP
and \Reddit, as they do not have node attributes.

\begin{table}[t]
	\vspace{-0.4cm}
	\caption{Profile of Datasets}
	\label{tab:dataset}
	\vspace{-0.2cm}
	\centering
	\tiny
	\resizebox{0.4\textwidth}{!}{
	\begin{tabular}{|c|r|r| r| r| r|}
		\hline
		\multicolumn{2}{|c|}{\textbf{Dataset}}   & \textbf{$|V(G)|$} &\textbf{ $|E(G)|$}  & \textbf{$ |\mathcal{A}| $} & \textbf{$| \mathcal{C}(G) |$} \\\hline 
		\multicolumn{2}{|c|}{\Cora}    & 2,708 & 5,429  & 1,433  &7\\         
		\multicolumn{2}{|c|}{\Citeseer} & 3,327 & 4,732 & 3,703 & 6\\
		\multicolumn{2}{|c|}{\Arxiv}&199,343 &1,166,243&N/A&40\\	
		\multicolumn{2}{|c|}{\DBLP} &317,080 &1,049,866  &N/A&5,000\\  
		\multicolumn{2}{|c|}{\Reddit}&232,965&114,615,892&N/A & 50 \\                                \hline
		\multirow{10}{*}{\begin{tabular}[c]{@{}c@{}}\Facebook \end{tabular}} 
		&0   & 348 & 2,867  & 224 & 24 \\ 
		&107 & 1,046    & 27,795  &  576   & 9  \\
		& 348 & 228    &3,420  &   162  &  14   \\
		& 414    & 160    & 1,853  & 105  & 7   \\
		& 686 & 171    &1,827  &   63   &  14  \\
		& 698 & 67    &337  &   48   &  13  \\
		& 1684 & 793    & 14,817  &   319 & 17\\
		& 1912 & 756    &30,781  &   480 & 46\\
		& 3437 & 548    &5,361  & 262  & 32\\
		& 3980 & 60    &206  &  42  & 17\\ \hline
	\end{tabular}
	}
	\vspace{-0.4cm}
\end{table}

\stitle{Tasks \& Queries:} We test our CGNP in different subgraphs of
same graph, different graphs, and different application scenarios,
%
following the 4 different types of tasks described in \cref{sec:problem}:
\ding{172} Single Graph Shared Communities Task (\SGSC), 
\ding{173} Single Graph Disjoint Communities Task (\SGDC), 
\ding{174} Multiple Graphs from One Domain Task (\MGOD), and   
\ding{175} Multiple Graphs from Different Domains Task (\MGDD).
%
For \SGSC, \SGDC and \MGDD, one task is generated by sampling a
subgraph of 200 nodes by BFS.
The query nodes are randomly drawn from a sampled subgraph,
$G$, where we assign 1 or 5 query nodes to the support set $\support$,
i.e., 1-shot or 5-shot tasks, and assign 30 query nodes to the query
set $\query$ disjointly.  It is worth noting that
%
%
the query nodes
%
%
may be from the same ground-truth communities for \SGSC whereas the
query nodes must be from disjoint communities for \SGDC.
For each query $q$, we randomly drawn 5 positive samples from the
community of $q$, $\mathcal{C}_{q}(G)$, to construct $l_q^{+}$ and 10
negative samples from $V(G) \setminus \mathcal{C}_{q}(G)$ to construct
$l_q^{-}$.
Here, for \SGSC and \SGDC, we
generate 100 training tasks for \Cora, \Citeseer, \Arxiv, \Reddit and
\DBLP, and generate 50 valid tasks and 50 test tasks for the five
datasets, respectively.
For \MGOD, we use one \Facebook ego-network as the graph in one task,
and sample the same numbers of queries and labels as discussed
above. Ten tasks are split into 6 for training, 2 for validation, and
2 for testing.
For \MGDD, we also generate 100 tasks of \Citeseer for training, 50
tasks of \Cora for validation, and 50 tasks of \Cora for testing,
denoted as \Citeseercora.

\stitle{Baselines:} To comprehensively evaluate the performance of
CGNP framework for CS, we compare with {10} baseline approaches,
including {3} graph algorithms, {4} naive approaches discussed in
\cref{sec:naive}, {3} traditional ML/DL-based approaches.
%
%
\ding{182} Attributed Truss Community Search (\ATC)~\cite{ATC}. It is
an attributed community search algorithm given the input of query
nodes and attributes. Firstly, it finds the maximal $(k, d)$-truss
containing the query nodes. Then, the algorithm iteratively removes
unpromising nodes from the truss, which has a small attribute score.
%
%
\ding{183} Attributed Community Query (\ACQ)~\cite{ACQ}. It aims to
find subgraph whose nodes are tightly connected and share common
attributes with the given query node.
%
%
\ding{184} Closest Truss Community (\CTC)~\cite{CTC}. It is a
$k$-truss based community search framework for non-attributed
graphs. Given a set of query nodes, $Q$, a greedy algorithm finds a
$k$-truss with the largest $k$ that contains $Q$ and has the minimum
diameter among the truss.
%
%
\ding{185} Model-Agnostic Meta-Learning (\MAML)~\cite{MAML}. We use 
GNN as the base model. The task-specific parameters of  GNN are
updated in an inner loop as Eq.~(\ref{eq:maml:inner}), and the
task-common parameters are updated in an outer loop as
Eq.~(\ref{eq:maml:outer}) over all training tasks.
\ding{186} First-Order Meta-Learning (\Reptile)~\cite{reptile}. As a
first-order alternative of \MAML, \Reptile adopts the same GNN as the
base model. Task-common parameters are updated by
Eq.~(\ref{eq:reptile:outer}) in an outer loop, over all the training
tasks.
\ding{187} Feature Transfer (\Featrans). A base GNN model is
pre-trained on all the training tasks. For a test task $\task^*=
(\support^*, \query^*)$, the final layer of the GNN is finetuned on
the support set $\support^*$ by one gradient step, while all the other
parameters are kept intact.
\ding{188} Graph Prototypical Network (\PN).  For each query $q$, 3
positive samples and 3 negative samples are randomly drawn from $l_q$
to compute the query-specific prototypes. We use Euclidean distance as
the distance function in Eq.~(\ref{eq:gpn:likelihood}).
\ding{189} Supervised GNN (\Supervise). One GNN model is trained for
each test task from scratch by the few-shot data in $\support^*$.
\ding{190} ICS-GNN (\ICSGNN) \cite{ICSGNN}. For each query node $q$, a
GNN model is trained by some positive and negative samples and
predicts a score for the remaining nodes.  Then, the algorithm finds a
subgraph connected to $q$, with a fixed number of nodes, aiming to
maximize the summation of the scores predicted by GNN.
%
\ding{191} AQD-GNN (\AQDGNN) \cite{AQDGNN}. The setting is similar
to \Supervise. For each test task, \AQDGNN trains the model from
scratch by the few-shot data in $\support^*$ and test in $\query^*$.
It is worth noting that \PN and \ICSGNN are different
from other learning-based approaches, where test query nodes are
required to have ground-truth. \PN uses the ground-truth to compute
the query-specific prototypes while \ICSGNN uses the ground-truth to
train a query-specific model.  These two approaches \emph{cannot fully
  generalize} to query nodes without any prior knowledge of
membership.

\stitle{Implementation and Settings:} 
We give the settings of 8 ML approaches, including our CGNP and 7 baselines, \MAML, \Reptile, \Featrans, \PN, \Supervise, \ICSGNN and \AQDGNN.
For the GNN encoder of CGNP and the base GNN models of the 6 baselines, the number of the GNN layers is 3, where each GNN layer has 128 hidden units and a Dropout probability of 0.2 by default.

We investigate popular GNN layers, including the vanilla Graph Convolutional Network (\GCN)~\cite{GCN}, Graph Attention Network (\GAT)~\cite{GAT} and \SAGE~\cite{SAGE}, and finally choose 
\GAT by default due to its high performance. 
For the MLP decoder of CGNP, we use a two-layer MLP with 512 hidden units. 
For the GNN decoder of CGNP, we use a two-layer GNN which has the same configuration as the encoder. 

The learning framework of CGNP and the 7 ML baselines are built on
PyTorch~\cite{pytorch} with PyTorch Geometric~\cite{torchgeo}.
We use Adam optimizer with a learning rate of $5 \times
  10^{-4}$ to train CGNP, \PN, \ICSGNN, \Supervise, and \Featrans by
  200 epochs.  For \MAML and \Reptile, the inner loop performs 10
  gradient steps for training and 20 steps for testing, with a
  learning rate of $5\times 10^{-4}$, and the learning rate for the
  outer loop is $10^{-3}$.
It is worth mentioning that the performance of CGNP is robust in the
range of empirical training hyper-parameters.  By default, the
training and prediction are conducted on a Tesla V100 with 16GB
memory.  \ATC, \ACQ and \CTC are tested on the same Linux server with
32 Intel(R) Silver 4,215 CPUs and 128GB RAM.

\stitle{Evaluation Metrics:} To evaluate the quality of the found result, we use accuracy, precision, recall and \Fone-score between the prediction and the ground-truth. 
\comment{Suppose $\hat{l} \in \{0, 1\}^{n}$ and $l \in \{0, 1\}^{n}$ are the binary representations of the prediction result and the ground-truth of a query $q$. Precision, recall and \Fone-score are defined as below:
\begin{align}
	\nonumber
	\Pre(\hat{l},l)&=\frac{\sum_{v \in V(G)} \hat{l}(v) \& l(v)}{\sum_{v \in V(G)}\hat{l}(v) }, 
	\Rec(\hat{l},l)=\frac{\sum_{v \in V(G)} \hat{l}(v) \& l(v)}{\sum_{v \in V(G)}{l(v)}}  \\
	\nonumber
	\Fone(\hat{l},l)&=\frac{2\cdot \Rec(\hat{l},l)\cdot \Pre(\hat{l},l)}{ \Rec(\hat{l},l) + \Pre(\hat{l},l)}
\end{align}
}
\Fone-score is the harmonic average of precision and recall, which better reflects the overall performance.
\begin{table*}[t]
	
	\centering
	\caption{Performance on \SGSC and \SGDC Tasks (First and Second Best \Fone Scores are Highlighted)}
	\vspace{-0.2cm}
	\label{tab:result1}
	\resizebox{1\textwidth}{!}{
		\begin{tabular}{|l|l|r|r|r|r|r|r|r|r|r|r|r|r|r|r|r|r|}
			\hline
			\multicolumn{1}{|c|}{\multirow{3}{*}{Dataset}} & \multicolumn{1}{c|}{\multirow{2}{*}{Task config.}} & \multicolumn{8}{c|}{Single Graph with Shared Communities}                                                                                                                                                               & \multicolumn{8}{c|}{Single Graph with Disjoint Communities}                                                                                                                                                            \\\cline{3-18}
			\multicolumn{1}{|c|}{}                         & \multicolumn{1}{c|}{}                                & \multicolumn{4}{c|}{1-shot}                                                                               & \multicolumn{4}{c|}{5-shot}                                                                              & \multicolumn{4}{c|}{1-shot}                                                                              & \multicolumn{4}{c|}{5-shot}                                                                              \\\cline{2-18}
			\multicolumn{1}{|c|}{}                         & \multicolumn{1}{c|}{Methods}                         & \multicolumn{1}{c|}{\Acc} & \multicolumn{1}{c|}{\Pre} & \multicolumn{1}{c|}{\Rec} & \multicolumn{1}{c|}{\Fone} & \multicolumn{1}{c|}{\Acc} & \multicolumn{1}{l|}{\Pre} & \multicolumn{1}{c|}{\Rec} & \multicolumn{1}{c|}{\Fone} & \multicolumn{1}{c|}{\Acc} & \multicolumn{1}{c|}{\Pre} & \multicolumn{1}{c|}{\Rec} & \multicolumn{1}{c|}{\Fone} & \multicolumn{1}{c|}{\Acc} & \multicolumn{1}{c|}{\Pre} & \multicolumn{1}{c|}{\Rec} & \multicolumn{1}{c|}{\Fone} \\\hline

			\multirow{11}{*}{\rotatebox{90}{\Citeseer}} 
			&\ATC       & 0.4759 & 0.8366 & 0.1044 & 0.1856 & 0.4623 & 0.8344 & 0.1005 & 0.1793 & 0.5393 & 0.8288 & 0.1131 & 0.1990 & 0.5373 & 0.8357 & 0.1144 & 0.2013 \\
			&\CTC       & 0.4386 & 0.8585 & 0.0226 & 0.0440 & 0.4264 & 0.8653 & 0.0225 & 0.0439 & 0.5043 & 0.8262 & 0.0262 & 0.0508 & 0.5010 & 0.8293 & 0.0261 & 0.0507 \\\cline{2-18}
			&\MAML      & 0.5293 & 0.6450 & 0.3942 & 0.4894 & 0.5494 & 0.6937 & 0.4108 & 0.5160 & 0.5528 & 0.5835 & 0.4071 & 0.4796 & 0.5738 & 0.6277 & 0.4022 & 0.4903 \\
			&\Reptile   & 0.5474 & 0.6382 & 0.4825 & 0.5495 & 0.5550 & 0.6886 & 0.4363 & 0.5342 & 0.5812 & 0.6038 & 0.5022 & 0.5483 & 0.5970 & 0.6500 & 0.4531 & 0.5340 \\
			&\Featrans  & 0.4719 & 0.6625 & 0.1571 & 0.2540 & 0.4548 & 0.6692 & 0.1337 & 0.2229 & 0.5044 & 0.5346 & 0.1602 & 0.2465 & 0.4925 & 0.5127 & 0.0819 & 0.1413 \\
			&\PN        & 0.1744 & 0.1159 & 0.1564 & 0.1332 & 0.1383 & 0.1208 & 0.1441 & 0.1314 & 0.4498 & 0.4632 & 0.6199 & 0.5302 & 0.2957 & 0.3960 & 0.3263 & 0.3578 \\\cline{2-18}
			&\Supervise & 0.5492 & 0.6574 & 0.4429 & 0.5293 & 0.5688 & 0.6895 & 0.4780 & 0.5646 & 0.5751 & 0.6072 & 0.4544 & 0.5198 & 0.6221 & 0.6692 & 0.5110 & 0.5795 \\
			&\ICSGNN    & 0.4856 & 0.7115 & 0.1783 & 0.2852 & 0.4793 & 0.7094 & 0.1760 & 0.2821 & 0.5424 & 0.6738 & 0.1925 & 0.2994 & 0.5411 & 0.6658 & 0.1905 & 0.2963 \\
			&\AQDGNN    &0.5036	&0.5993	&0.4406	&0.5079  &0.5263	&0.5806	&0.6816	&0.6270 &0.5558	&0.5717	&0.5291	&0.5496 &0.5072	&0.5100	&0.7761	&0.6155\\\cline{2-18}
			&\CGNPIP    & 0.6076 & 0.6429 & 0.7071 & \cellcolor{LightCyan}{0.6734} & 0.6150 & 0.6562 & 0.7176 & \cellcolor{LightCyan}{0.6855} & 0.5611 & 0.5488 & 0.7469 & \cellcolor{LightCyan}{0.6327} & 0.5626 & 0.5515 & 0.7584 & \cellcolor{LightCyan}{0.6386} \\
			&\CGNPMLP   & 0.6041 & 0.6556 & 0.6490 & 0.6523 & 0.6160 & 0.6710 & 0.6735 & 0.6723 & 0.5510 & 0.5427 & 0.7174 & 0.6179 & 0.5773 & 0.5633 & 0.7588 & \cellcolor{LightRed}{0.6466} \\
			&\CGNPGNN   & 0.6133 & 0.6393 & 0.7443 & \cellcolor{LightRed}{0.6878} &  0.6158 & 0.6513 & 0.7367 &\cellcolor{LightRed} {0.6914} & 0.5685 & 0.5527 & 0.7730 & \cellcolor{LightRed}{0.6446} & 0.5765 & 0.5631 & 0.7532 & 0.6444\\\hline\hline
			
			\multirow{11}{*}{\rotatebox{90}{\Arxiv}} 
			&\ATC       & 0.5802 & 0.7253 & 0.0734 & 0.1333 & 0.5850 & 0.7349 & 0.0757 & 0.1373 & 0.5767 & 0.7875 & 0.0542 & 0.1015 & 0.5804 & 0.7930 & 0.0557 & 0.1042 \\
			&\CTC       & 0.5751 & 0.7693 & 0.0484 & 0.0911 & 0.5795 & 0.7783 & 0.0501 & 0.0942 & 0.5733 & 0.8160 & 0.0411 & 0.0782 & 0.5766 & 0.8200 & 0.0415 & 0.0790 \\\cline{2-18}
			&\MAML      & 0.5674 & 0.6512 & 0.0355 & 0.0673 & 0.5903 & 0.8005 & 0.0806 & 0.1465 & 0.5692 & 0.7345 & 0.0355 & 0.0676 & 0.5770 & 0.7221 & 0.0544 & 0.1011 \\
			&\Reptile   & 0.5762 & 0.6409 & 0.0829 & 0.1468 & 0.5888 & 0.8260 & 0.0726 & 0.1334 & 0.5697 & 0.6034 & 0.0693 & 0.1242 & 0.5719 & 0.6804 & 0.0409 & 0.0771 \\
			&\Featrans  & 0.5735 & 0.6527 & 0.0647 & 0.1177 & 0.5762 & 0.6626 & 0.0577 & 0.1062 & 0.5744 & 0.7288 & 0.0546 & 0.1016 & 0.5775 & 0.7066 & 0.0589 & 0.1087 \\
			&\PN        & 0.2588 & 0.2458 & 0.2061 & 0.2242 & 0.2754 & 0.2773 & 0.2453 & 0.2603 & 0.4681 & 0.6257 & 0.5354 & \cellcolor{LightCyan}0.5771 & 0.4195 & 0.4867 & 0.6055 & 0.5397 \\\cline{2-18}
			&\Supervise & 0.5531 & 0.4742 & 0.1488 & 0.2265 & 0.5816 & 0.6512 & 0.0877 & 0.1545 & 0.5461 & 0.4508 & 0.1364 & 0.2094 & 0.5791 & 0.6245 & 0.0957 & 0.1659 \\
			&\ICSGNN    & 0.5904 & 0.6004 & 0.2016 & 0.3019 & 0.5896 & 0.5995 & 0.2011 & 0.3012 & 0.5968 & 0.6224 & 0.2153 & 0.3199 & 0.5999 & 0.6220 & 0.2168 & 0.3215 \\
			&\AQDGNN    & 0.5183 & 0.4622 & 0.5217 & 0.4901 & 0.4821 & 0.4425 & 0.7266 & 0.5501 & 0.5215 & 0.4573 & 0.5105 & 0.4824 & 0.5069 & 0.4619 & 0.7220 & 0.5633 \\\cline{2-18}
			&\CGNPIP    & 0.5172 & 0.4716 & 0.8118 & \cellcolor{LightCyan}{0.5966} & 0.5520 & 0.4915 & 0.7889 & \cellcolor{LightRed}{0.6057} & 0.5699 & 0.5076 & 0.8067 & \cellcolor{LightRed}{0.6231} & 0.5856 & 0.5170 & 0.8083 & \cellcolor{LightRed}{0.6306} \\
			&\CGNPMLP   & 0.5079 & 0.4649 & 0.7870 & 0.5845 & 0.5642 & 0.5003 & 0.7371 & 0.5960 & 0.5365 & 0.4806 & 0.6397 & 0.5489 & 0.5847 & 0.5194 & 0.6841 & 0.5905 \\
			&\CGNPGNN   & 0.4699 & 0.4496 & 0.9161 & \cellcolor{LightRed}{0.6032} &  0.4649 & 0.4449 & 0.9205 & \cellcolor{LightCyan}{0.5998} & 0.4938 & 0.4548 & 0.7464 & 0.5652  &0.5520	&0.4950	&0.8399	&\cellcolor{LightCyan}{0.6229}\\\hline\hline

			\multirow{11}{*}{\rotatebox{90}{\Reddit}} 
			&\ATC       & 0.6574 & 0.3566 & 0.4286 & 0.3893 & 0.6582 & 0.3553 & 0.4282 & 0.3883 & 0.4784 & 0.9586 & 0.4108 & 0.5752 & 0.4787 & 0.9572 & 0.4136 & 0.5776 \\
			&\CTC       & 0.6614 & 0.3593 & 0.4202 & 0.3874 & 0.6627 & 0.3583 & 0.4190 & 0.3863 & 0.4713 & 0.9593 & 0.4019 & 0.5664 & 0.4722 & 0.9577 & 0.4054 & 0.5697 \\\cline{2-18}
			&\MAML      & 0.7450 & 0.3254 & 0.0007 & 0.0014 & 0.7465 & 0.3812 & 0.0010 & 0.0020 & 0.4679 & 0.9864 & 0.3861 & 0.5550 & 0.5017 & 0.9863 & 0.4277 & 0.5967 \\
			&\Reptile   & 0.7414 & 0.3039 & 0.0116 & 0.0224 & 0.7447 & 0.2874 & 0.0050 & 0.0098 & 0.4051 & 0.9904 & 0.3107 & 0.4730 & 0.4046 & 0.9907 & 0.3121 & 0.4746 \\
			&\Featrans  & 0.7327 & 0.2972 & 0.0361 & 0.0644 & 0.7393 & 0.3224 & 0.0261 & 0.0484 & 0.2784 & 0.9369 & 0.1719 & 0.2906 & 0.2345 & 0.8634 & 0.1328 & 0.2302 \\
			&\PN        & 0.4731 & 0.2285 & 0.5556 & 0.3238 & 0.5051 & 0.2253 & 0.5379 & 0.3175 & 0.6708 &0.9891 &0.6749 &0.8024 & 0.6622 &0.9871 &0.6675 &0.7965 \\\cline{2-18}
			&\Supervise & 0.7079 & 0.2677 & 0.0845 & 0.1284 & 0.7288 & 0.2546 & 0.0364 & 0.0637 & 0.5834 & 0.9736 & 0.5296 & 0.6860 & 0.5536 & 0.9827 & 0.4907 & 0.6545 \\
			&\ICSGNN    & 0.6949 & 0.3331 & 0.1959 & 0.2467 & 0.6975 & 0.3361 & 0.1990 & 0.2500 & 0.2748 & 0.9460 & 0.1652 & 0.2813 & 0.2725 & 0.9499 & 0.1652 & 0.2815 \\
			&\AQDGNN    & 0.5221 & 0.2514 & 0.4427 & 0.3207 & 0.4141 & 0.2616 & 0.7199 & 0.3837 & 0.6476 & 0.8851 & 0.6772 & 0.7673 & 0.7830 & 0.9139 & 0.8250 & 0.8672 \\\cline{2-18}
			&\CGNPIP    & 0.2557 & 0.2549 & 0.9991 & \cellcolor{LightRed}{0.4062} & 0.2543 & 0.2534 & 0.9983 & 0.4042 & 0.7885 & 0.9264 & 0.8184 & 0.8691 & 0.8482 & 0.9110 & 0.9122 & 0.9116 \\
			&\CGNPMLP   & 0.2566 & 0.2544 & 0.9934 & 0.4051 & 0.2774 & 0.2557 & 0.9694 & \cellcolor{LightRed}{0.4047} & 0.8229 & 0.9397 & 0.8479 & \cellcolor{LightCyan}{0.8915} & 0.8697 & 0.9508 & 0.8945 & \cellcolor{LightRed}{0.9218} \\
			&\CGNPGNN   & 0.2548 & 0.2548 & 1.0000 & \cellcolor{LightCyan}{0.4061} & 0.2534 & 0.2534 & 1.0000 & \cellcolor{LightCyan}{0.4043} & 0.8578 & 0.8578 & 1.0000 & \cellcolor{LightRed}{0.9235} & 0.8584 & 0.8584 & 1.0000 & \cellcolor{LightCyan}{0.9238}\\\hline\hline
			
			\multirow{11}{*}{\rotatebox{90}{\DBLP}} 
			&\ATC       & 0.8376 & 0.8749 & 0.1752 & 0.2919 & 0.8230 & 0.8916 & 0.1676 & 0.2822 & 0.7527 & 0.7539 & 0.0922 & 0.1643 & 0.7360 & 0.7849 & 0.1038 & 0.1834 \\
			&\CTC       & 0.8365 & 0.9107 & 0.1599 & 0.2720 & 0.8216 & 0.9214 & 0.1534 & 0.2629 & 0.7512 & 0.7711 & 0.0803 & 0.1454 & 0.7345 & 0.8012 & 0.0931 & 0.1668 \\\cline{2-18}
			&\MAML      & 0.8161 & 0.6395 & 0.0864 & 0.1522 & 0.8029 & 0.6545 & 0.1065 & 0.1832 & 0.7383 & 0.5337 & 0.0581 & 0.1047 & 0.7201 & 0.5713 & 0.0776 & 0.1366 \\
			&\Reptile   & 0.8106 & 0.5135 & 0.1704 & 0.2559 & 0.7993 & 0.5833 & 0.1162 & 0.1938 & 0.7208 & 0.3890 & 0.1033 & 0.1632 & 0.7184 & 0.5508 & 0.0741 & 0.1306 \\
			&\Featrans  & 0.8194 & 0.6339 & 0.1296 & 0.2152 & 0.8057 & 0.6600 & 0.1315 & 0.2193 & 0.7417 & 0.5736 & 0.0796 & 0.1397 & 0.7238 & 0.6301 & 0.0789 & 0.1402 \\
			&\PN        & 0.1819 & 0.0120 & 0.4528 & 0.0235 & 0.1790 & 0.0748 & 0.6846 & 0.1349 & 0.4017 & 0.2292 & 0.5911 & 0.3303 & 0.3581 & 0.2408 & 0.3988 & 0.3003 \\\cline{2-18}
			&\Supervise & 0.7312 & 0.2142 & 0.1523 & 0.1780 & 0.7773 & 0.3987 & 0.1438 & 0.2113 & 0.6805 & 0.3075 & 0.1692 & 0.2183 & 0.7015 & 0.4255 & 0.1307 & 0.2000 \\
			&\ICSGNN    & 0.7997 & 0.4662 & 0.3571 & \cellcolor{LightRed}{0.4044} & 0.7911 & 0.5030 & 0.3519 & \cellcolor{LightCyan}{0.4141} & 0.7366 & 0.4978 & 0.2304 & 0.3150 & 0.7290 & 0.5414 & 0.2373 & 0.3299 \\
			&\AQDGNN    & 0.6129 & 0.2257 & 0.4220 & 0.2941 & 0.5615 & 0.2705 & 0.6556 & 0.3830 & 0.5421 & 0.2990 & 0.5737 & 0.3931 & 0.4567 & 0.2994 & 0.6992 & 0.4192 \\\cline{2-18}
			&\CGNPIP    & 0.3951 & 0.2206 & 0.8548 & \cellcolor{LightCyan}{0.3507} & 0.5320 & 0.2829 & 0.8175 & \cellcolor{LightRed}{0.4203} & 0.3988 & 0.2779 & 0.8288 & \cellcolor{LightRed}{0.4162} & 0.5166 & 0.3404 & 0.7720 & \cellcolor{LightCyan}{0.4725} \\
			&\CGNPMLP   & 0.4926 & 0.2317 & 0.7147 & 0.3499 & 0.5203 & 0.2487 & 0.6488 & 0.3596 & 0.4437 & 0.2814 & 0.7415 & \cellcolor{LightCyan}{0.4080} & 0.5652 & 0.3632 & 0.7304 & \cellcolor{LightRed}{0.4851} \\
			&\CGNPGNN   & 0.4262 & 0.2223 & 0.8018 & 0.3481 & 0.4535 & 0.2463 & 0.7928 & 0.3759 & 0.3999 & 0.2682 & 0.7644 & 0.3971  & 0.3777 & 0.2894 & 0.8377 & 0.4302
			\\\hline
		\end{tabular}
	}
\end{table*}

\subsection{Effectiveness}
\label{sec:exp:effect}

We investigate the overall performance of CGNP on the four types of
tasks (\SGSC, \SGDC, \MGOD and \MGDD), for 1-shot and 5-shot
learning. The number of shots is the number of query nodes provided in
the support set.  The three variants of CGNP, CGNP with simple inner
product decoder (\CGNPIP), CGNP with MLP decoder (\CGNPMLP), and CGNP
with GNN decoder (\CGNPGNN) are compared with 10 baseline approaches.

Table \ref{tab:result1} presents the performance for tasks of single
graph with shared/disjoint communities. Here, we
highlight the first (purple) and the second (blue) best \Fone.
We observe that CGNP outperforms all the baselines in most cases.
The \Fone of CGNP succeeds all the baselines 0.28 on average.
The superiority of CGNP is reflected in improving the recall
significantly, while keeping relatively high accuracy and precision.
In the testing, we observe that the optimization-based approaches, e.g., MAML, Reptile, predict almost all the nodes as the negative samples.  These approaches are sensitive to the imbalanced label distribution, leading to a higher accuracy but low recall. 
That indicates accuracy is not a suitable metric to evaluate the overall performance for CS task, because most nodes are in the negative class. A model is easy to achieve high accuracy as long as it predicts more nodes as the negative samples.
\ICSGNN performs best in some cases (e.g., \DBLP and \Facebook), as it
is a query-specific model and uses the ground-truth of the test query
nodes additionally.  The naive approaches like \Featrans even fail to
search the community in most cases due to their low \Fone score. As
for ML/DL-based methods, they get comparable scores with naive
approaches. Since these methods are trained from scratch for each new
test task or each new query, ML/DL-based methods can utilize
task-specific knowledge. However, our approach gets higher score than
these methods. It indicates that our method is an efficient
meta-learning approach, which can learn prior knowledge from different
tasks. The prior knowledge is beneficial for predicting the
community. CGNP is the most robust learner due to its metric-based
learning strategy, and this property is similar to KNN and GP, which
fully validates the effectiveness of CGNP for small data.

	\begin{table}[t]
	\centering
	\caption{Performance on \MGOD and \MGDD Tasks}
	\vspace{-0.3cm}	
	\label{tab:result2}
	\resizebox{0.5\textwidth}{!}{
		\begin{tabular}{|l|l|r|r|r|r|r|r|r|r|}
			\hline
			\multirow{2}{*}{Dataset}        & \multicolumn{1}{l|}{Task config.} & \multicolumn{4}{c|}{1-shot}                                                                             & \multicolumn{4}{c|}{5-shot}                                                                             \\ \cline{2-10} 
			& \multicolumn{1}{c|}{Methods}        & \multicolumn{1}{c|}{\Acc} & \multicolumn{1}{c|}{\Pre} & \multicolumn{1}{c|}{\Rec} & \multicolumn{1}{c|}{\Fone} & \multicolumn{1}{c|}{\Acc} & \multicolumn{1}{c|}{\Pre} & \multicolumn{1}{c|}{\Rec} & \multicolumn{1}{c|}{\Fone} \\ \hline
			\multirow{12}{*}{\rotatebox{90}{\Facebook}}       
			&\ATC       & 0.5564 & 0.2595 & 0.6305 & 0.3677 & 0.5592 & 0.2611 & 0.6464 & 0.3720 \\
			&\ACQ       & 0.3625 & 0.2190 & 0.8248 & 0.3461 & 0.4109 & 0.2266 & 0.7944 & 0.3526 \\
			&\CTC       & 0.8518 & 0.8734 & 0.3224 & 0.4710 & 0.8540 & 0.8904 & 0.3159 & 0.4664 \\\cline{2-10}
			&\MAML      & 0.6050 & 0.2319 & 0.1692 & 0.1956 & 0.6806 & 0.4091 & 0.2687 & 0.3244 \\
			&\Reptile   & 0.6356 & 0.3049 & 0.2215 & 0.2566 & 0.6680 & 0.4251 & 0.4642 & 0.4438 \\
			&\Featrans  & 0.6105 & 0.2462 & 0.1804 & 0.2082 & 0.5867 & 0.2936 & 0.3192 & 0.3059 \\
			&\PN        & 0.1549 & 0.0648 & 0.1289 & 0.0863 & 0.0938 & 0.0469 & 0.0967 & 0.0631 \\\cline{2-10}
			&\Supervise & 0.6291 & 0.2343 & 0.1350 & 0.1713 & 0.6073 & 0.3421 & 0.4079 & 0.3721 \\
			&\ICSGNN    & 0.7606 & 0.5722 & 0.5598 & \cellcolor{LightRed}{0.5659} & 0.7574 & 0.5906 & 0.5516 & \cellcolor{LightRed}{0.5704} \\
			&\AQDGNN    & 0.6779 & 0.3548 & 0.1644 & 0.2247 & 0.4239 & 0.3112 & 0.8396 & 0.4540\\\cline{2-10}
			&\CGNPIP    & 0.3727 & 0.3107 & 0.9925 & {0.4733} & 0.5121 & 0.3666 & 0.9756 & {0.5329} \\
			&\CGNPMLP   & 0.4161 & 0.3203 & 0.9418 &\cellcolor{LightCyan} {0.4781} & 0.5659 & 0.3860 & 0.8832 &{0.5372} \\
			&\CGNPGNN   & 0.3077 & 0.2888 & 0.9835 & 0.4465 & 0.6029	&0.4118	&0.9145	&\cellcolor{LightCyan}{0.5678}\\\hline\hline
			
			\comment{
				\multirow{11}{*}{\rotatebox{90}{\Coraciteseer}} 
				&\ATC       & 0.5421 & 0.8159 & 0.0726 & 0.1333 & 0.5308 & 0.8123 & 0.0676 & 0.1249 \\
				&\CTC       & 0.5231 & 0.8455 & 0.0209 & 0.0407 & 0.5135 & 0.8474 & 0.0207 & 0.0405 \\\cline{2-10}
				&\MAML      & 0.5132 & 0.4979 & 0.4111 & 0.4504 & 0.5271 & 0.5378 & 0.3164 & 0.3984 \\
				&\Reptile   & 0.5282 & 0.5191 & 0.3737 & 0.4346 & 0.5452 & 0.5593 & 0.3823 & 0.4542 \\
				&\Featrans  & 0.5108 & 0.4843 & 0.1293 & 0.2041 & 0.5119 & 0.5213 & 0.1676 & 0.2537 \\
				&\PN        & 0.2983 & 0.3553 & 0.3309 & 0.3427 & 0.3185 & 0.4131 & 0.3936 & 0.4031 \\\cline{2-10}
				&\Supervise & 0.5331 & 0.5280 & 0.3549 & 0.4245 & 0.5506 & 0.5607 & 0.4246 & 0.4833 \\
				&\ICSGNN    & 0.5564 & 0.6708 & 0.1622 & 0.2613 & 0.5492 & 0.6800 & 0.1613 & 0.2608 \\\cline{2-10}
				&\CGNPIP    & 0.5495 & 0.5429 & 0.8674 & \cellcolor{LightCyan}{0.6678} & 0.5475 & 0.5396 & 0.9013 & \cellcolor{LightRed}{0.6750} \\
				&\CGNPMLP   & 0.5464 & 0.5403 & 0.8789 & \cellcolor{LightRed}{0.6692} & 0.5516 & 0.5487 & 0.7876 & 0.6468 \\
				&\CGNPGNN   & 0.5544 & 0.5529 & 0.7649 & 0.6419 & 0.5427 & 0.5398 & 0.8324 & \cellcolor{LightCyan}{0.6549}   \\\hline\hline
			}
			
			\multirow{11}{*}{\rotatebox{90}{\Citeseercora}} 
			&\ATC       & 0.5779 & 0.8191 & 0.1885 & 0.3064 & 0.5783 & 0.8154 & 0.1934 & 0.3127 \\
			&\CTC       & 0.5166 & 0.8714 & 0.0269 & 0.0523 & 0.5156 & 0.8692 & 0.0271 & 0.0526 \\\cline{2-10}
			&\MAML      & 0.5042 & 0.4986 & 0.3630 & 0.4202 & 0.5334 & 0.5544 & 0.3008 & 0.3900 \\
			&\Reptile   & 0.5202 & 0.5219 & 0.3620 & 0.4275 & 0.5658 & 0.6066 & 0.3538 & 0.4469 \\
			&\Featrans  & 0.5289 & 0.5529 & 0.2498 & 0.3442 & 0.5122 & 0.5464 & 0.0960 & 0.1632 \\
			&\PN        & 0.2521 & 0.1716 & 0.3800 & 0.2364 & 0.1766 & 0.1656 & 0.3524 & 0.2254 \\\cline{2-10}
			&\Supervise & 0.5446 & 0.5537 & 0.4099 & 0.4711 & 0.6066 & 0.6206 & 0.5320 & 0.5729 \\
			&\ICSGNN    & 0.5532 & 0.6642 & 0.1923 & 0.2982 & 0.5538 & 0.6677 & 0.1932 & 0.2996 \\
			&\AQDGNN    & 0.5145 & 0.5040 & 0.5685 & 0.5343 & 0.4652 & 0.4626 & 0.6365 & 0.5358\\\cline{2-10}
			&\CGNPIP    & 0.5351 & 0.5177 & 0.8822 & 0.6525 & 0.5280 & 0.5134 & 0.9241 & \cellcolor{LightRed}0.6601 \\
			&\CGNPMLP   & 0.5397 & 0.5207 & 0.8781 & \cellcolor{LightCyan}{0.6537} & 0.5476 & 0.5267 & 0.8654 & \cellcolor{LightCyan}{0.6548} \\
			&\CGNPGNN   & 0.5367 & 0.5179 & 0.9181 & \cellcolor{LightRed}{0.6623} &  0.5456 & 0.5191 & 0.8532 & 0.6455\\\hline
		\end{tabular}
	}
\end{table}

Table \ref{tab:result2} shows the performance for tasks of multiple graphs.
The tasks of multiple graphs are harder than that of the single graph, and the tasks across domains are even harder. 
The \Fone of CGNP surpasses the \Fone of all the baselines 0.25 by average. 
The CGNP variants dominate the top two best models on \Citeseercora while it is overwhelmed by \ICSGNN on \Facebook.
This demonstrates that CGNP can effectively learn prior knowledge from only a few data of one graph and adapt to other graphs even from different domains, and the learned prior is indeed helpful.
In fact, transferring the prior of a shared node embedding function for clustering, as what CGNP does, is much easier than transferring model parameters, as \MAML, {\Reptile} and \Featrans do. 

CGNP with different decoders may bring different performance to the
result. The difference between them is subtle i.e. less than $5\%$. 
Both \MAML and \Reptile perform worse than CGNP in general.  The graph
algorithms \ATC, \ACQ and \CTC also fail to outperform learning-based
approaches due to their low recall. 
It is worth to mentioning that \ACQ fails to return the results for \Citeseercora and \Citeseer in 12 hours, since \ACQ needs to enumerate all the sets of attributes that are shared by the query node and candidates.
In addition, \ACQ relies on the node attributes and it cannot support graphs
without node attributes, such as \Arxiv, \DBLP and \Reddit.

\subsection{Efficiency}
\label{sec:exp:effic}

\begin{figure}[t]
	\centering
	\begin{tabular}[h]{c}		
		\subfigure[{Total Test Time}] {\label{fig:test time}
			\includegraphics[ width=0.85\columnwidth]{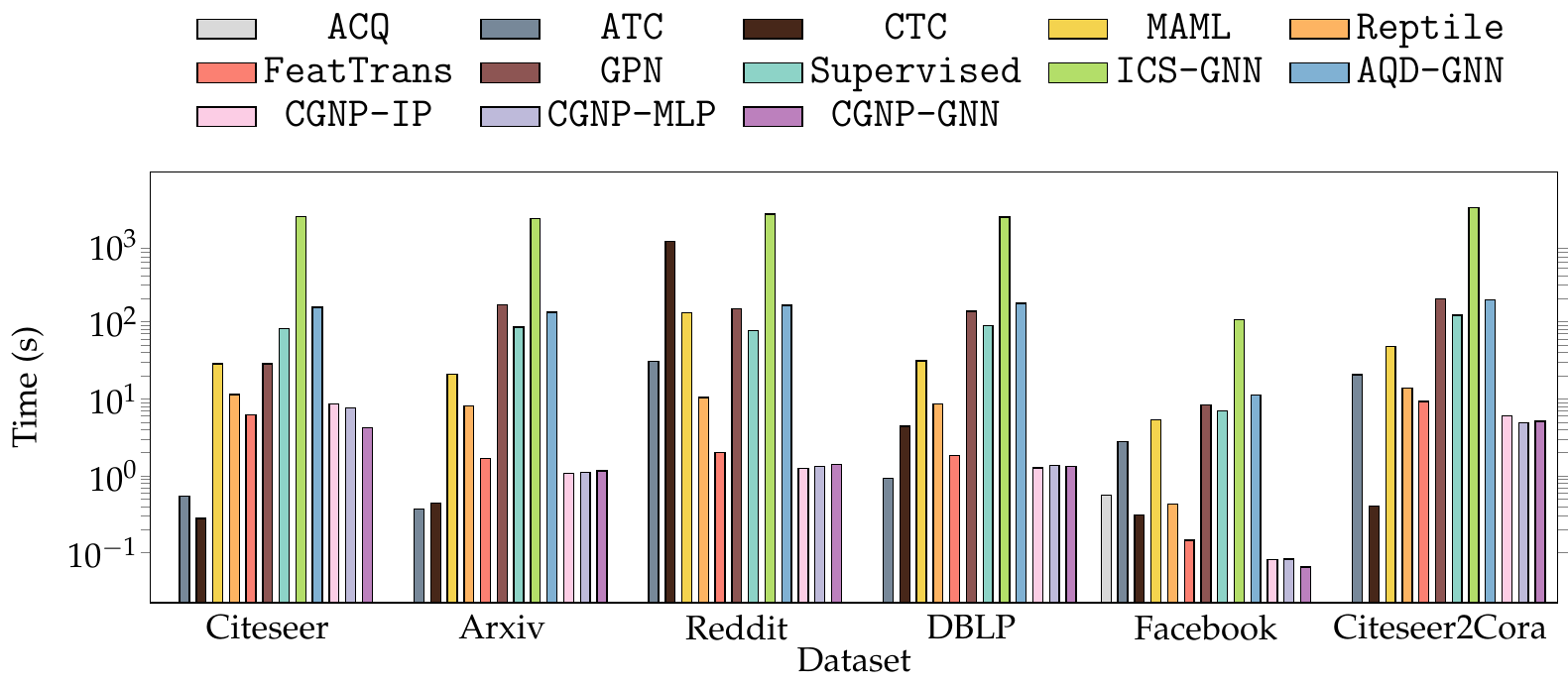}
		} \\
		\subfigure[{Total Training Time}] {\label{fig:train time}
			\includegraphics[ width=0.85\columnwidth]{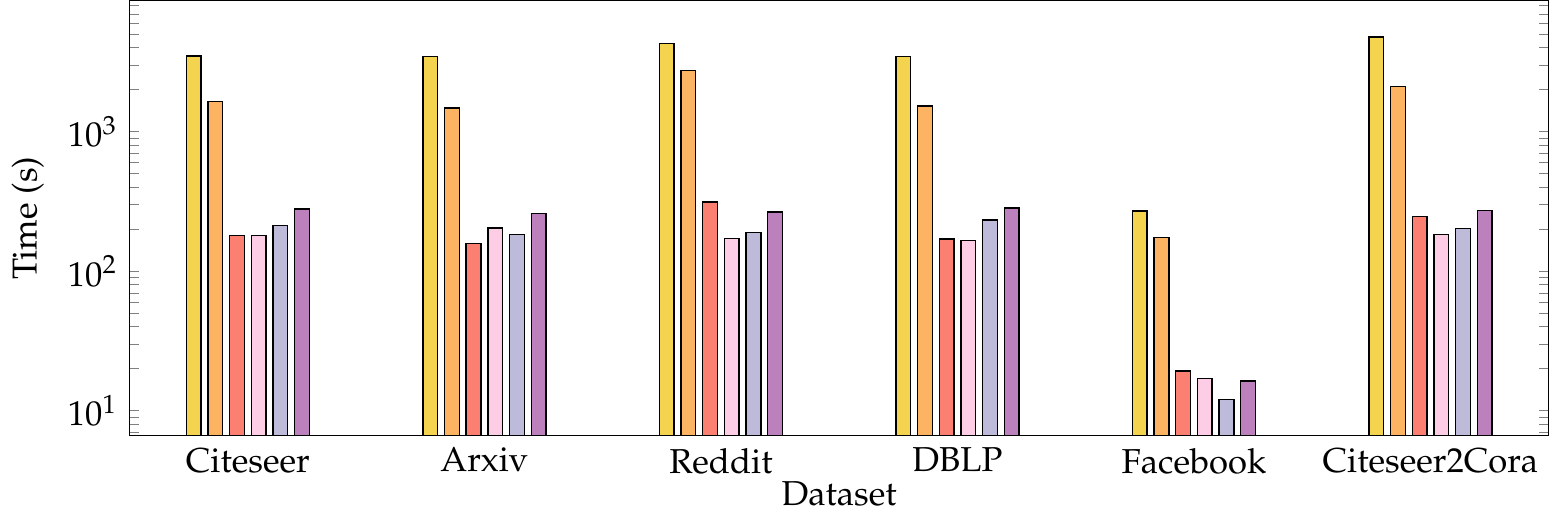}
		}
	\end{tabular}
	\vspace{-0.2cm}
	\caption{Training \& Test Time (s)}
	\vspace{-0.2cm}
	\label{fig:time}
\end{figure}

We compare the efficiency of CGNP and the baselines regarding the
test/training time.  Fig.~\ref{fig:test time} presents the total test
time.  Regarding the prediction efficiency, our CGNP is the best
learning-based approach and the second-best among all the approaches,
which is over one order of magnitude faster than \ATC, \ACQ, \MAML,
\Reptile, \PN, \Supervise, \ICSGNN and \AQDGNN, and slightly faster
than \Featrans.  For one test task, \MAML, \Reptile and \Featrans
apply the backward propagation algorithm to update the parameters
online, and \Supervise and \AQDGNN train the parameters from scratch.
\ICSGNN needs to train a model for each query node on-the-fly.  \PN
not only needs to apply the backward propagation algorithm to update
parameters but also has to compute the distance between each node and
prototypes.  \ACQ needs to enumerate all the sets of
attributes shared by the query node and candidates, so that
it fails to return the results for \Citeseercora and \Citeseer in 12
hours.  For \CTC, the intermediate candidate communities of \Reddit
and \Facebook are large, it takes much longer time to compute the
diameter and maintain the $k$-truss structure.

Fig.~\ref{fig:train time} shows the meta training time of the
learning-based approaches on the training task set, where all the
models are trained by the same epoch of 200.  Note that \ATC, \ACQ,
\CTC, \PN, \Supervise, \ICSGNN and \AQDGNN do not involve this meta
training stage.  Our CGNP is one order of magnitude faster than \MAML
and \Reptile and its training efficiency is close to the simplest
transfer strategy, \Featrans,  \MAML and \Reptile are quite
time-consuming due to their two-level optimization paradigm.  For the
three CGNP variants, due to different model complexities, training
\CGNPGNN is slightly slower than that of \CGNPMLP, which is further
slightly slower than that of \CGNPIP. We observe that these
differences are negligible in the testing stage in Fig.~\ref{fig:test
  time}.

\begin{figure}[t]
	\centering
	\includegraphics[width=0.75\columnwidth]{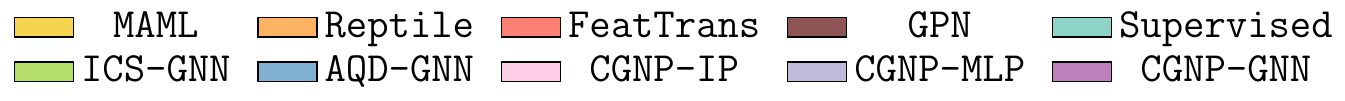}
	\begin{tabular}[h]{c}
		\hspace{-0.6cm}
		\subfigure[{Total Test Time}] {\label{fig:scalable test time}
			\includegraphics[ width=0.48\columnwidth]{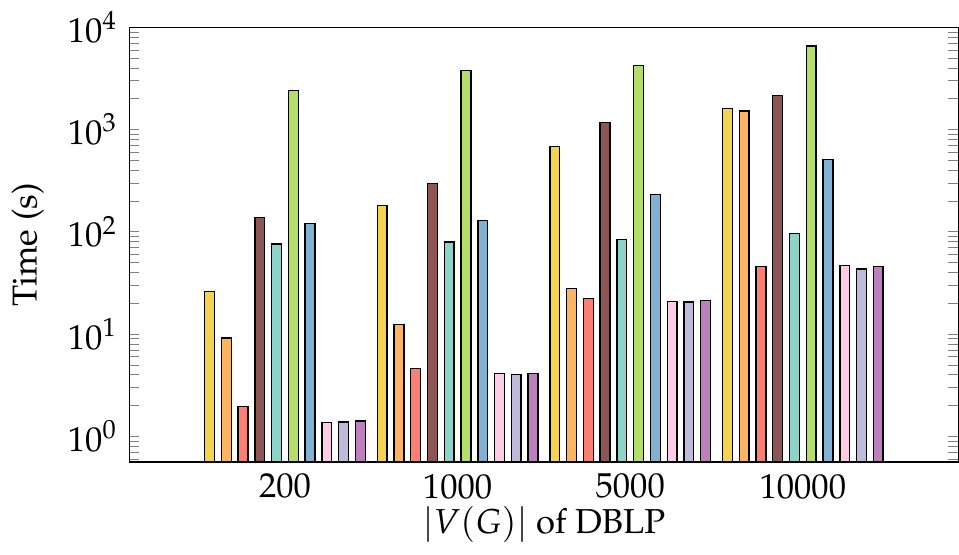}
		} 
		\hspace{-0.2cm}
		\subfigure[{Total Training Time}] {\label{fig:scalable train time}
			\includegraphics[ width=0.48\columnwidth]{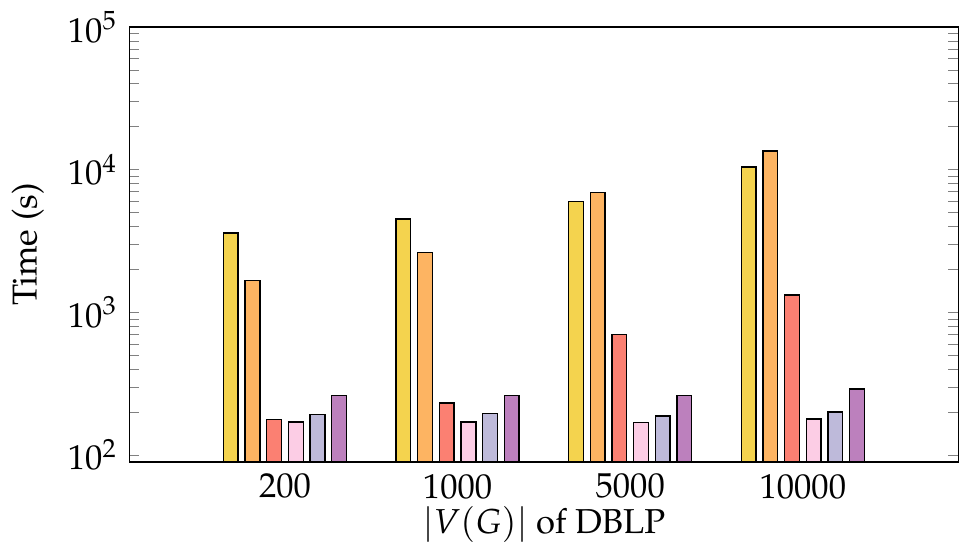}
		}
	\end{tabular}
	\vspace{-0.2cm}
	\caption{Scalability of Training \& Test (s)}
	\vspace{-0.6cm}
	\label{fig:scalabletime}
\end{figure}

\stitle{Scalability Test.}  We explore the scalability of the
learning-based approaches. Fig.~\ref{fig:scalabletime} shows the GPU
training time and test time of our CGNP and 6 ML baselines, as the
number of nodes of graph in each task increases from $1,000$ to
$10,000$.
%
%
All the methods can scale to graphs of $10,000$ nodes in the limited 16GB GPU memory.
%
%
The test time of CGNP costs the least time than other baselines in all
the sizes. Only \Featrans spends close time to CGNP. Other methods
like \MAML, \Reptile, \PN, \AQDGNN show similar results as shown in
Fig.~\ref{fig:test time} due to their training strategy.
%
%
While the training time of CGNP does not increase significantly with
the size increasing. And it is one or two orders of magnitude faster
than other methods in large graph.

\subsection{Effect of the ground-truth number}
\label{sec:exp:labels}

\begin{figure}[t]
	\vspace{-0.2cm}
	\centering
	\includegraphics[width=0.8\columnwidth]{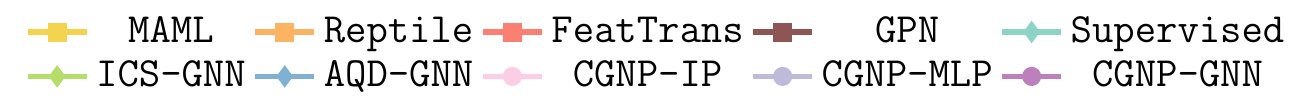}
	\begin{tabular}[h]{c}
		\hspace{-0.6cm}
		\subfigure[\Citeseer ] {
			\includegraphics[ width=0.4\columnwidth]{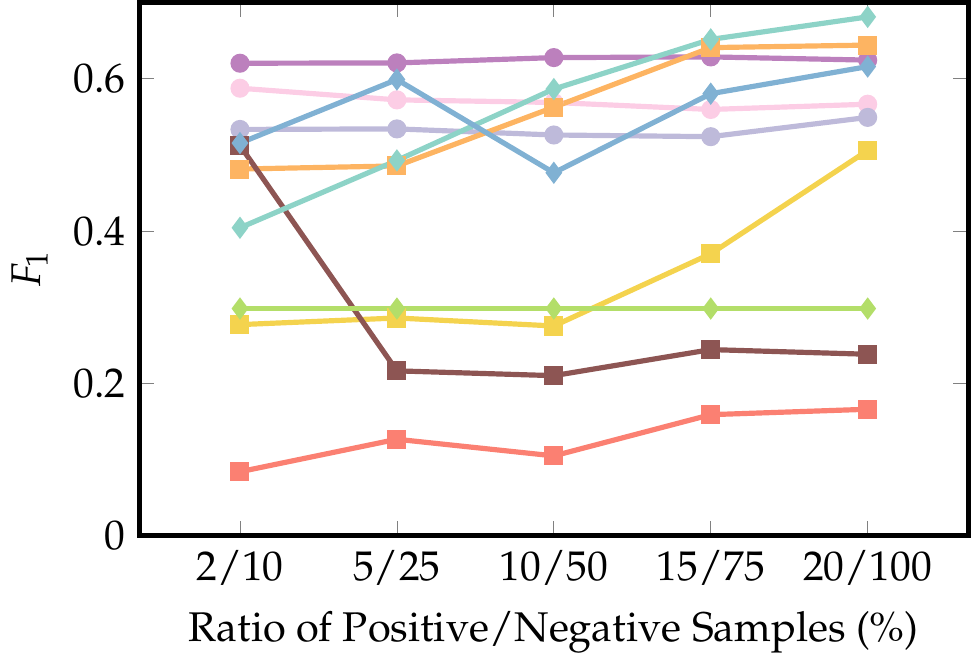}
			\label{fig:maskciteseer}
		}
		\subfigure[\Arxiv] {
			\includegraphics[ width=0.4\columnwidth]{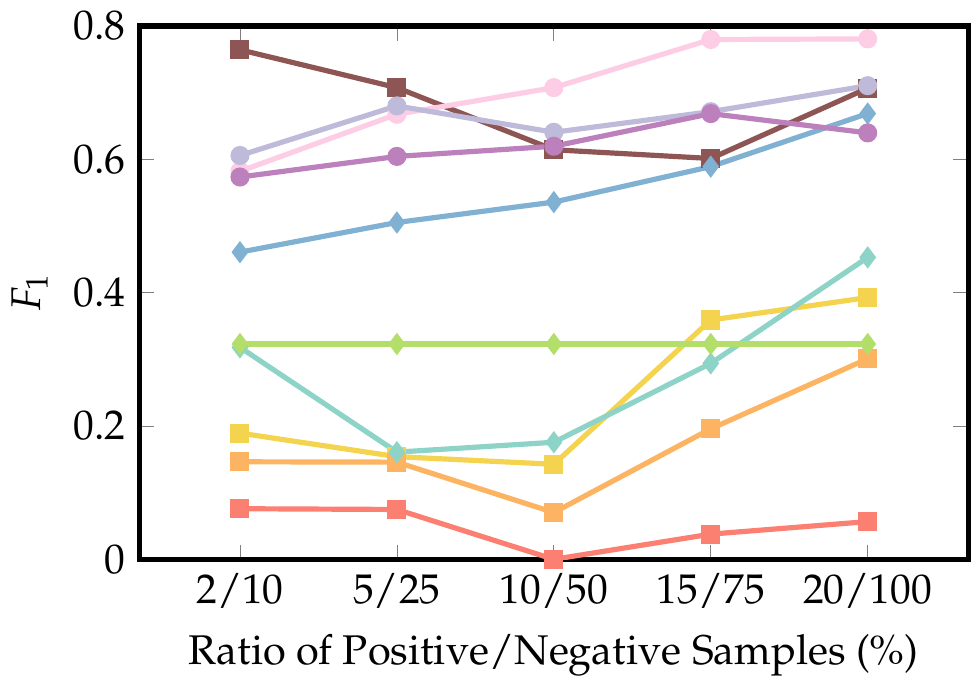}
			\label{fig:maskarxiv}
		}\\ 
		
		\hspace{-0.6cm}
		\subfigure[\Reddit] {
			\includegraphics[ width=0.4\columnwidth]{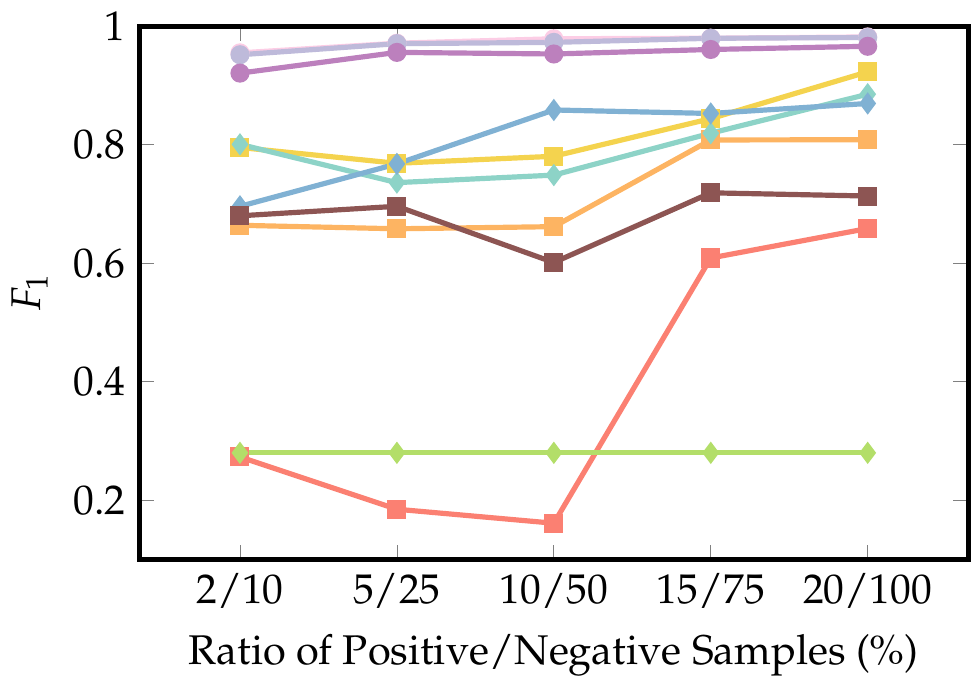}
			\label{fig:maskreddit}
		}
		\subfigure[\DBLP ] {
			\includegraphics[ width=0.4\columnwidth]{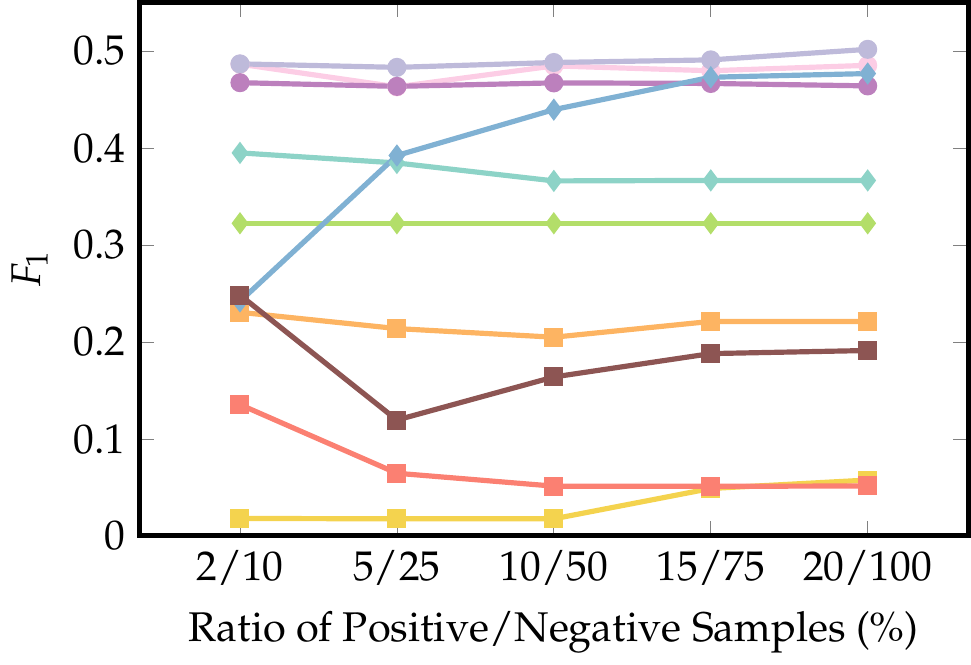}
			\label{fig:maskdblp}
		}\\
		
		\hspace{-0.6cm}
		\subfigure[\Facebook] {
			\includegraphics[ width=0.4\columnwidth]{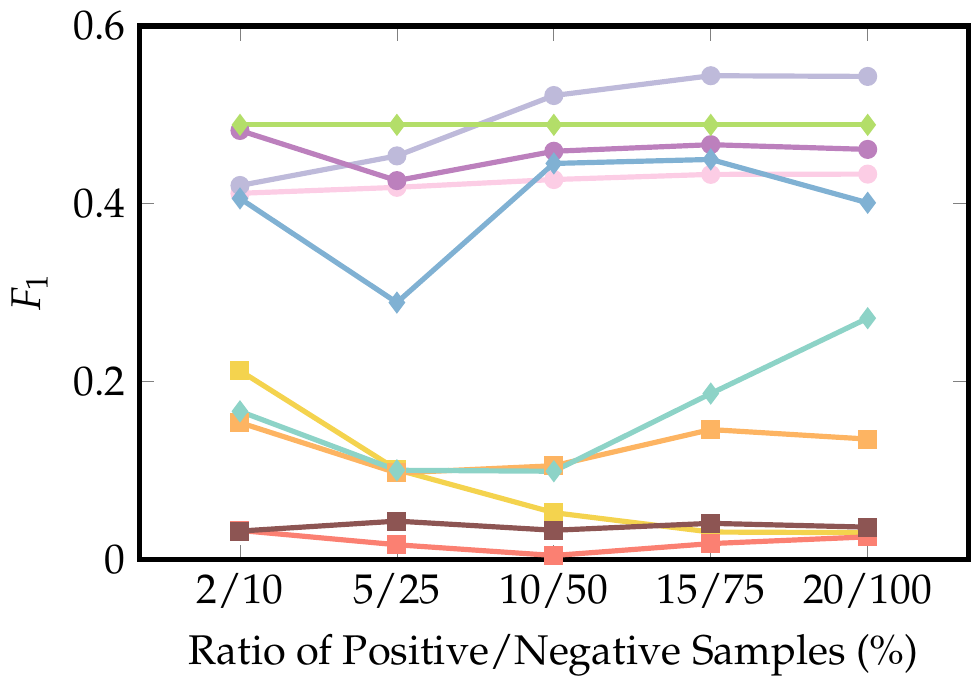}
			\label{fig:maskfacebook}
		}
		\subfigure[\Citeseercora] {
			\includegraphics[ width=0.4\columnwidth]{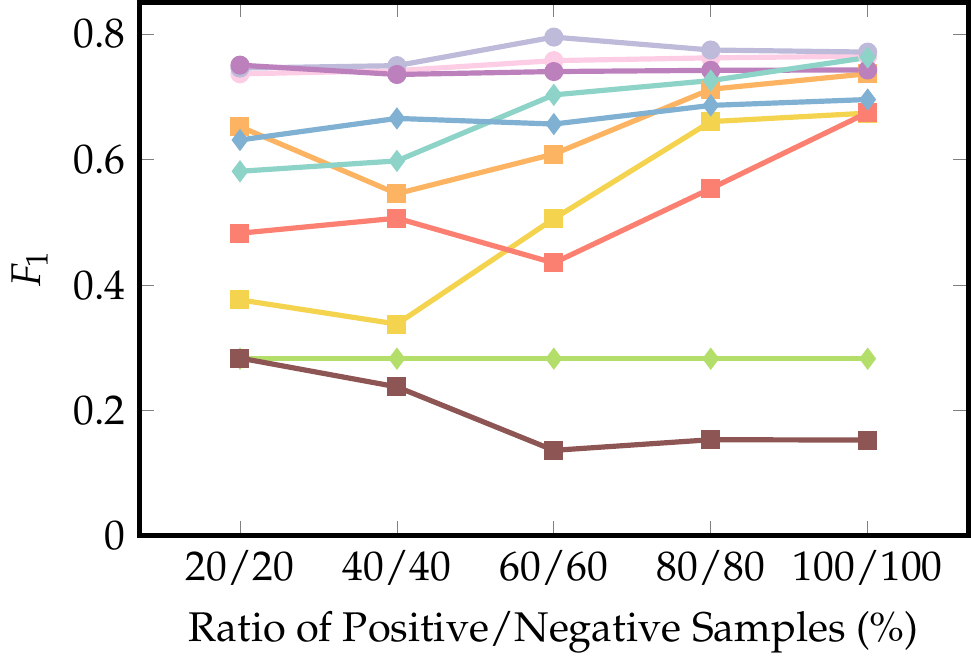}
			\label{fig:maskcoraciteseer}
		}\\
	\end{tabular}
	\vspace{-0.2cm}
	\caption{\Fone under Different Ratios of Ground-truth}
	\label{fig:mask}
	\vspace{-0.6cm}
\end{figure}

We further evaluate how the number of ground-truth samples influences
the performance of the learning-based approaches.
%
For each query node $q$ in the support set, we vary the number of
positive/negative samples, i.e., $|l_q^{+}|$/$|l_q^{-}|$,
from 2\%/10\% to 20\%/100\% of the total number of the nodes.
Fig.~\ref{fig:mask} shows the \Fone-score of the 3 CGNP variants and 7
ML baselines on the 6 different tasks in 1-shot scenario.  In
Fig.~\ref{fig:mask}, CGNP variants surpass the ML baselines by 30\% on
average, particularly under the circumstances of the scarce
ground-truth. For small training samples, \Supervise suffers from
severe over-fitting, and \Featrans and \PN also face a high risk of
over-fitting in their adaptation step.  As the number of ground-truth
increases, the performance of \PN would degrade, since more samples
may blur the representation of prototypes.  The performance of \MAML, \Reptile and \AQDGNN increase in general with the increasing number of ground
truth. In addition, \Supervise would overtake CGNP as shown in
Fig.~\ref{fig:maskciteseer}, when the number of ground-truth is at a
high level.  Given sufficient training data, a task-specific
\Supervise model can better adapt to the task, compared with a
meta model.
%
The \Fone of \ICSGNN tends to be stable on varying ratios of
samples. We conjecture the reason could be its hyper-parameter, the
community size to search, determines the final results in the
post-processing stage.
Furthermore, we find that the performance of CGNP is robust to the
number of ground-truth. That is in accordance with the nature of
metric-based learning, where only a few training samples can achieve
high performance for KNN and kernel learning.


\subsection{Ablation Study}
\label{sec:exp:ablation}

In this section, we conduct ablation studies to investigate the effect
of different options for the GNN layer and the commutative operation
on the performance of CGNP.  \CGNPMLP, the CGNP with an GNN decoder,
is tested as the base CGNP model and all the model variants are
trained by the same hyper-parameters.
These model variants are tested on the $5$-shot \Citeseer~\SGSC,
\Arxiv~\SGSC, \Reddit~\SGDC, \DBLP~\SGDC, \Facebook~\MGOD and
\Citeseercora~\MGDD tasks, respectively.

\stitle{GNN Layer.} We adopt three popular GNN, \GCN~\cite{GCN},
\GAT \cite{GAT} and \SAGE \cite{SAGE} as the encoder,
where the commutative operation is fixed to the average
pooling. Table~\ref{tab:layer} lists the performance of the \CGNPGNN
variants on the 2 tasks.
%
In general, the \GAT encoder consistently outperforms \GCN
encoders. This is because \GAT aggregates the node representation
weighted by learnable weights via self-attention, where the importance
of each neighbor are considered regarding its local structure,
possible features and positive/negative labels. The higher \Fone of
\GAT demonstrates the attention mechanism can also contribute to
improving the performance of CGNP in the encoder part. \SAGE encoder
gets highest \Fone scores in \SGDC and \SGSC tasks. This is because
\SAGE uses a generalized aggregation function and this mechanism is
beneficial for encoder of CGNP.

\stitle{Commutative Operation.} We adopt the sum, average pooling and
self-attention, introduced in \cref{sec:CGNP} as the commutative
operation big $\oplus$ of \CGNPGNN, by fixing \GAT as the encoder GNN.
%
Table~\ref{tab:layer} shows the corresponding performance of the 3
model variants. In different tasks, the performance of three
commutative operations is different.  However, the differences between
the three variants are relatively slight.  We speculate that different
tasks, graphs or ground-truth distributions may benefit from different
commutative operations, and the effect of the type of commutative
operation is not as remarkable as that of the GNN encoder.

\begin{table}[t]
	\small
	\centering
	\vspace{-0.4cm}
	\caption{Performance with Different Layers and Com. Op.}
	\vspace{-0.2cm}
	\label{tab:layer}
	\resizebox{0.5\textwidth}{!}{
		\begin{tabular}{|l|l|c|c|c|c|l|c|c|c|c|}
			\hline
			Dataset                   & \multicolumn{1}{c|}{Layer} & \multicolumn{1}{c|}{\Acc}    &  \multicolumn{1}{c|}{\Pre}    &  \multicolumn{1}{c|}{\Rec}    &  \multicolumn{1}{c|}{\Fone}  & \multicolumn{1}{c|}{$\oplus$}  & \multicolumn{1}{c|}{\Acc}    &  \multicolumn{1}{c|}{\Pre}    &  \multicolumn{1}{c|}{\Rec}    &  \multicolumn{1}{c|}{\Fone}   \\\hline
			\multirow{3}{*}{\Citeseer}    & GCN   &0.5001	&0.4601	&0.7779	&0.5782 & Att. & 0.5956 & 0.6437 & 0.6893 & 0.6657 \\
			& GAT   & 0.5520	&0.4950	&0.8399	&0.6229 & Sum       & 0.6154 & 0.6526 & 0.7306 & 0.6894 \\
			& SAGE  &0.6348	&0.5555	&0.8553	&\cellcolor{LightYellow}{0.6736}  & Ave.   & 0.6158 & 0.6513 & 0.7367 &\cellcolor{LightYellow} {0.6914} \\\hline
			\multirow{3}{*}{\Arxiv}  & GCN   & 0.4540 & 0.4388 & 0.9076 & 0.5916 & Att. & 0.4816 & 0.4526 & 0.9080 & \cellcolor{LightYellow}{0.6041}\\
			& GAT   & 0.4649 & 0.4449 & 0.9205 & 0.5998 & Sum       & 0.4696 & 0.4405 & 0.8044 & 0.5692 \\
			& SAGE  & 0.6035 & 0.5305 & 0.7800 & \cellcolor{LightYellow}{0.6315} 	& Ave.   & 0.4649 & 0.4449 & 0.9205 & 0.5998 \\\hline
			\multirow{3}{*}{\Reddit} & GCN   & 0.8006 & 0.8596 & 0.9175 & 0.8876 & Att.  & 0.8006 &0.8006 &1.0000 &0.8893\\
			& GAT   & 0.8584 & 0.8584 & 1.0000 & 0.9238 & Sum       &0.7122 &0.7122 &1.0000 &0.8319\\
			& SAGE  & 0.9335 & 0.9553 & 0.9679 & \cellcolor{LightYellow}{0.9615} & Ave.   & 0.8584 & 0.8584 & 1.0000 & \cellcolor{LightYellow}{0.9238} \\\hline
			\multirow{3}{*}{\DBLP}   & GCN   & 0.3189 & 0.2829 & 0.9308 & 0.4339  & Att. & 0.3761 & 0.2866 & 0.8223 & 0.4251 \\
			& GAT   & 0.3777 & 0.2894 & 0.8377 & 0.4302 & Sum       & 0.3742 & 0.2896 & 0.8472 & \cellcolor{LightYellow}{0.4316} \\
			& SAGE  & 0.5480 & 0.3446 & 0.6783 & \cellcolor{LightYellow}{0.4570} & Ave.   & 0.3777 & 0.2894 & 0.8377 & 0.4302 \\\hline
			\multirow{3}{*}{\Facebook} & GCN   & 0.3082	&0.2880	&0.9676	&0.4438 & Att. & 0.5547	&0.3795	&0.8826	&0.5307 \\
			& GAT   & 0.6029	&0.4118	&0.9145	&\cellcolor{LightYellow}{0.5678} & Sum       & 0.5427	&0.3762	&0.9156	&0.5333 \\
			& SAGE  &0.4361	&0.3143	&0.8263	&0.4554  & Ave.   & 0.6029	&0.4118	&0.9145	&\cellcolor{LightYellow}{0.5678} \\\hline
			\multirow{3}{*}{\Citeseercora} & GCN   & 0.5427 & 0.5179 & 0.8224 & 0.6356 & Att. & 0.5626 & 0.5310 & 0.8390 & \cellcolor{LightYellow}{0.6504} \\
			& GAT   & 0.5456 & 0.5191 & 0.8532 & \cellcolor{LightYellow}{0.6455} & Sum       & 0.5341 & 0.5113 & 0.8898 & 0.6494 \\
			& SAGE  & 0.5591 & 0.5306 & 0.7867 & 0.6337 & Ave. 	& 0.5456 & 0.5191 & 0.8532 & 0.6455\\\hline
		\end{tabular}
	}
\end{table}